\documentclass[lettersize,journal]{IEEEtran}
\usepackage{amsmath,amsfonts}
\usepackage{algorithmic}
\usepackage{algorithm}
\usepackage{array}
\usepackage{color}
\usepackage{subfig}
\usepackage{bm}
\usepackage{textcomp}
\usepackage{stfloats}
\usepackage{url}
\usepackage{verbatim}
\usepackage{graphicx}
\usepackage{epstopdf}
\usepackage{cite}
\usepackage[colorlinks=true,linkcolor=black,citecolor=black,urlcolor=black]{hyperref}
\usepackage[table,svgnames]{xcolor}

\usepackage{float}
\usepackage{multicol}
\usepackage{multirow}
\usepackage{tikz-imagelabels}
\usepackage{tikz}
\usepackage{pict2e}
\usepackage{overpic}
\usepackage{booktabs}
\imagelabelset{
	coordinate label font = \large,
	coordinate label back = white,
	coordinate label text = black,
}
\newcommand{\round}[1]{\ensuremath{\lfloor#1\rceil}}
\newcommand{\RNum}[1]{\uppercase\expandafter{\romannumeral #1\relax}}

\begin{document}

\title{Semantic-aided Parallel Image Transmission Compatible with Practical System}

\author{Mingkai Xu, Yongpeng Wu, \IEEEmembership{Senior Member,~IEEE},\thanks{M. Xu, Y. Wu (\emph{corresponding author}) and W. Zhang are with the Department of Electronic Engineering, Shanghai Jiao Tong University, Shanghai 200240, China (e-mail: {michaelxu, yongpeng.wu, zhangwenjun}@sjtu.edu.cn).} Yuxuan Shi,\thanks{Y. Shi is with the School of Cyber and Engineering, Shanghai Jiao Tong University, Shanghai 200240, China (e-mail: ge49fuy@sjtu.edu.cn).} Xiang-Gen Xia, \IEEEmembership{Fellow,~IEEE},\thanks{X.-G. Xia is with the Department of Electrical and Computer Engineering, University of Delaware, Newark, DE 19716, USA (e-mail: xxia@ee.udel.edu).} \\M{\'e}rouane Debbah, \IEEEmembership{Fellow,~IEEE},\thanks{M. Debbah is with KU 6G Research Center, Khalifa University of Science and Technology, P O Box 127788, Abu Dhabi, UAE (email: merouane.debbah@ku.ac.ae) and also with CentraleSupelec, University Paris-Saclay, 91192 Gif-sur-Yvette, France.} Wenjun Zhang, \IEEEmembership{Fellow,~IEEE}, Ping Zhang, \IEEEmembership{Fellow,~IEEE}\thanks{P. Zhang is with the State Key Laboratory of Networking and Switching Technology, Beijing University of Posts and Telecommunications, Beijing 100876, China (e-mail: pzhang@bupt.edu.cn).}}

\maketitle

\begin{abstract}
		In this paper, we propose a novel semantic-aided image communication framework for supporting the compatibility with practical separation-based coding architectures. Particularly, the deep learning (DL)-based joint source-channel coding (JSCC) is integrated into the classical separate source-channel coding (SSCC) to transmit the images via the combination of semantic stream and image stream from DL networks and SSCC respectively, which we name as parallel-stream transmission. The positive coding gain stems from the sophisticated design of the JSCC encoder, which leverages the residual information neglected by the SSCC to enhance the learnable image features. Furthermore, a conditional rate adaptation mechanism is introduced to adjust the transmission rate of semantic stream according to residual, rendering the framework more flexible and efficient to bandwidth allocation. We also design a dynamic stream aggregation strategy at the receiver, which provides the composite framework with more robustness to signal-to-noise ratio (SNR) fluctuations in wireless systems compared to a single conventional codec. Finally, the proposed framework is verified to surpass the performance of both traditional and DL-based competitors in a large range of scenarios and meanwhile, maintains lightweight in terms of the transmission and computational complexity of semantic stream, which exhibits the potential to be applied in real systems.
	\end{abstract}
	

	\section{Introduction}
	\IEEEPARstart{F}OR several decades, the conventional communication paradigm has been advanced under the guidance of Shannon information theory, which aims for effective and reliable transmission with bit-level accuracy. To guarantee this, state-of-the-art communication systems perform source coding and channel coding separately, then modulating the bits into constellation points for transmission. Take image communication as an example. An advanced compression format (e.g., JPEG~\cite{jpeg}, JPEG2000~\cite{jpeg2000}, BPG~\cite{bpg}) combined with a capacity-achieving channel code (e.g., low-density parity check (LDPC)~\cite{ldpc} code, Turbo~\cite{5075875} code) is widely implemented in a practical system with efficient hardware deployment. Such a separate source-channel coding (SSCC) design, however, yields suboptimal performance in the common finite-blocklength setting. Moreover, conventional image transmission experiences a drastic performance loss when the channel condition deteriorates, known as the 'cliff effect'.
	
	This hence motivates the consideration of a novel communication paradigm named semantic communication, which prefers restoring the meanings of transmitted information in terms of specific tasks rather than reconstructing the bits accurately \cite{Burks_Shannon_Weaver_1951}. Since only the task-related semantic information is extracted from the source, the amount of transmitted data is much less than that of original data when pursuing similar task performances, which largely reduces the communication cost. Furthermore,
	deep learning (DL) has sparked great interest in communication research~\cite{huang2024compression,liu2022time,10574376,9083671} including but not limited to effective compression, reliable transmission, and device-edge-cloud collaboration, where a deep adaptive inference network is introduced to provide accurate and low-latency performance in resource-constrained scenarios~\cite{10261454}. In the context of semantic communication, joint source-channel coding (JSCC)~\cite{Fresia2010} design based on DL methods is powerful in extracting high-level semantic features from the input data, which drives the framework to achieve end-to-end optimality. More recently, there have emerged DL-based semantic communication frameworks for multi-modal data~\cite{10158995,Wang2023,xie2021,xiao2023,Huang2022,10093953}, which adopt trainable deep neural networks (DNN) to extract and transmit the semantic information of images, videos, texts, speech, point clouds and time series, in order to execute the downstream tasks at the receiver.

	Nevertheless, there are inevitably some limitations in DL-driven JSCC that could potentially hinder its application.
	\subsubsection{Flexibility} Flexible control of the transmission rate is crucial for the deployment of communication frameworks. In the SSCC scheme, such control can be effortless achieved by separately adjusting the source, channel coding methods. However, for most DL networks, the transmission rate remains invariant until the networks being retrained, which causes inconvenience in coping with dynamic channel bandwidth requirements. To address this issue, the authors of \cite{djscc-l} transmit multiple layers in a progressive manner for reconstruction improvement; In~\cite{dynamicjscc}, features are divided into groups where the transmitted portion is determined by source and channel distributions; In~\cite{ntscc}, a nonlinear transform source-channel coding method is proposed to adapt the overall rate by a learnable entropy model. However, these rate adaptation mechanisms are inflexible in practice, which is due to the very limited number of rate values provided or the fact that the overall transmission rate is unforeseeable until the model inference stage.
	
	\subsubsection{Performance} The DL-based semantic communication framework for image is prone to performance saturation with the increase of source dimension~\cite{ntscc}. This is mainly attributed to the fact that the highly-integrated artificial neural networks (ANN) in semantic coding modules hardly provide sufficient model expression capability for large-scale images. To tackle this shortcoming, some recent works~\cite{witt,9830752,ntscc} employ Transformer~\cite{transformer} as the backbone of the transceiver and succeed in outperforming conventional SSCC schemes in terms of task performances. However, this superiority comes at the cost of model complexity, leading to high computation cost, memory burden, and processing latency. In addition, an intrinsic deficiency of end-to-end DL lies in its optimization on a specific dataset, which implies that the learned parameters of the JSCC are tailored to the source distributions at training. Once the distributions shift, the trained JSCC is likely to experience a decline in performance or even fail to transmit the source data. This phenomenon can be interpreted as the unavailability of a universal semantic-level JSCC~\cite{qin2021semantic}. In contrast, this flaw is avoidable in conventional SSCC with highly-generalized compression algorithms and capacity-achieving channel coding strategies.

    Consequently, a direct idea is to couple a learnable JSCC model into an SSCC system so that a DL-based semantic transceiver and a conventional image codec can complement each other's weaknesses. In this work, we propose a parallel-stream wireless semantic communication framework (\textbf{ParaSC}), in which two types of streams are transmitted over the channel: a non-trainable \emph{image stream} from the conventional SSCC encoder, and a learnable \emph{semantic stream} from the DL-driven JSCC transmitter. More specifically, the working principle of ParaSC is as follows: the conventional image encoder first compresses an image into bits, which are subsequently protected by a channel encoder and modulated to pass through the channel, and then decoded as a distorted image. Simultaneously, the convolutional neural network (CNN)-based JSCC encoder extracts the semantic information via residual coding, which utilizes the discrepancy between the original and decompressed image to effectively compensate for compression loss~\cite{Huang20212}. Beyond that, we exploit this residual to guide the rate adaptation of semantic features, thereby reducing the communication overhead of the semantic stream. At the receiver, the JSCC decoder dynamically integrates semantic information with the decoded distorted image to ensure reliable semantic recovery. Equipped with the semantic stream, on one hand, the original SSCC scheme becomes more robust in low SNR environment; on the other hand, the semantic stream is produced by a lightweight JSCC encoder, which hardly increases the complexity of the whole framework and thus enhances the compatibility with practical systems. Besides, the overall transmission rate of ParaSC is widely tunable and predictable, implying a great flexibility of the framework. In summary, we present the contributions in this paper as follows:

	\begin{itemize}
		\item We propose ParaSC, a semantic-enhanced communication framework that achieves the transmission and integration of image and semantic streams. In this framework, the DL-based JSCC assists the classical SSCC by incorporating the learnable semantics into the received image. Comparatively, our CNN-based JSCC codec has lower model complexity and bandwidth usage than other DL-based counterparts, which enables this framework to enjoy the efficient hardware implementation for conventional coding~\cite{nguyen2019efficient} and be compatible with practical communication architectures.

		\item At the transmitter, by leveraging the residual which contains the information not involved in SSCC, we realize the interaction between SSCC and JSCC to compensate for the compression loss from conventional coding and enhance the extracted semantic representations. At the receiver, the residual-enhanced semantic stream and the conventional image stream are merged by the semantic decoder, in which a parallel aggregation network (PAGNet) is proposed to aggregate the two streams with dynamic weights for combating the channel noise.
		
		\item We provide a flexible rate adaptation mechanism within ParaSC in terms of the residual. We adopt a conditional entropy model for the learnable stream to formulate the optimization function via an evidence lower bound (ELBO). By training the rate-adaptive codec, the semantic stream will be allocated proper resources according to its entropy conditioned on residual representations, thereby saving the communication overhead without compromising the performance. 
		
		\item The performance of ParaSC is verified in comparison with conventional separation-based coding schemes and DL-based frameworks for wireless transmission. ParaSC outperforms these baseline schemes in a large range of channel environments and bandwidth scenarios in terms of various quality assessment metrics. Ablation studies verify that the semantic stream remains lightweight while effectively mitigating the notorious 'cliff effect' associated with the conventional coding scheme. This demonstrates that our framework is compatible and complementary to the currently-deployed image coding systems.
\end{itemize}

The paper is organized as follows. In Section \ref{Sec2}, we first introduce the wireless image communication model and propose the ParaSC framework. In Section \ref{Sec3}, the DL-based JSCC codec is introduced. In Section \ref{Sec4}, the optimization function of ParaSC is derived theoretically and the approach of conditional semantic rate adaptation is presented. Furthermore, the numerical results are shown in Section \ref{Sec5}. Finally, Section \ref{Sec6} concludes the paper.

	\section{System Model and Introduction of ParaSC}
	\label{Sec2}
	
	In this section, we first describe the notational convenience. We then establish the system model for point-to-point wireless image communication and provide a general introduction of the proposed ParaSC framework.
	
	\subsection{Notational Convenience}
	Throughout the paper, we denote scalars and vectors by lowercase letters in normal and bold fonts, respectively, e.g., scalar $x$ and vector $\bm{x}=(x_1,x_2,\cdots,x_L)^T$. The alphabet of the scalar $x$ is defined as the corresponding calligraphy letter as $x\in\mathcal{X}$ and $|\mathcal{X}|$ represents the cardinality of a set $\mathcal{X}$. Moreover, $\mathbb{R}$ and $\mathbb{C}$ denote the sets of real and complex numbers, respectively. For a complex vector $\bm{z}\in\mathbb{C}^L$, $\bm{z}^{*}$ denotes its conjugate transpose. The identity matrix of size $k$ is denoted as $\boldsymbol{I}_k$, while $\boldsymbol{A}^{-1}$ and $\boldsymbol{A}^H$ are the inverse and Hermitian of a matrix $\boldsymbol{A}$, respectively. We define the distribution of an $n$-dimensional complex Gaussian random variable as $\mathcal{CN}(\bm{\mu},\bm{\Sigma})$ with mean $\bm{\mu}$ and covariance matrix $\bm{\Sigma}$. The uniform distribution of a univariate random variable on interval $[a,b]$ is denoted as $\mathcal{U}(a,b)$.
	\begin{figure*}[tbp]
	\centering
	\includegraphics[width=0.9\textwidth]{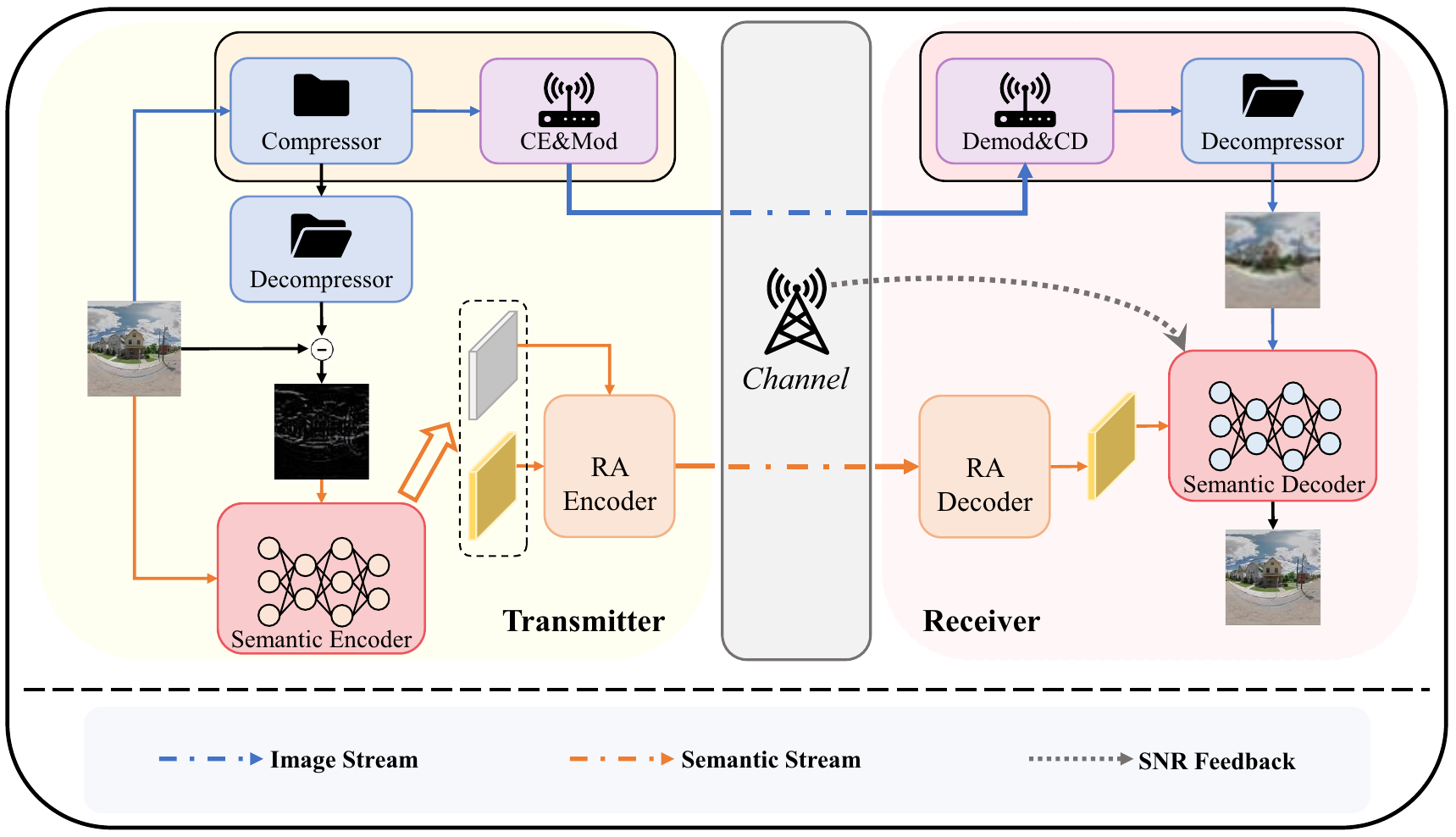}
  \begin{picture}(100,0)
  	{\large
  	\put(-123,169){$\bm{x}$}
  	\put(-108,137){$\bm{x}_r$}
  	\put(-76,171){$\bm{x}_c$}
  	\put(-23,144){$\bm{r}$}
  	\put(-23,107){$\bm{s}$}
  	\put(35,215){${\bm{z}}_w$}
  	\put(35,134){${\bm{z}}_s$}
    \put(100,215){$\hat{\bm{z}}_w$}
  	\put(100,134){$\hat{\bm{z}}_s$}
  	\put(174,118){$\hat{\bm{s}}$}
  	\put(196,193){$\hat{\bm{x}}_c$}
  	\put(222,68){$\hat{\bm{x}}$}
  	\small
  	\put(-49,259){$F_{\theta}{(\cdot)}$}
  	\put(-93,109){$F_{\varphi}{(\cdot,\cdot)}$}
  	\put(-1,140){$F_{\psi}{(\cdot,\cdot)}$}
  	\put(123,140){$G_{\xi}{(\cdot)}$}
  	\put(208,158){$G_{\zeta}{(\cdot,\cdot,\cdot)}$}
  	\footnotesize
  	\put(172,259){$F^{-1}_{\theta}(\cdot)$}
  	}
\end{picture}
	\caption{An overview of the ParaSC framework. The blue line stands for the conventional branch while the orange line represents the semantic branch, with the transmitted streams denoted by dash-dotted lines. The grey dotted line shows the SNR feedback link. "CE\&Mod", "Demod\&CD", "RA Encoder" and "RA Decoder" represent the channel encoding plus modulation, demodulation plus channel decoding, rate-adaptive encoder, and rate-adaptive decoder, respectively.}
	\label{fig:framework}
	\end{figure*}
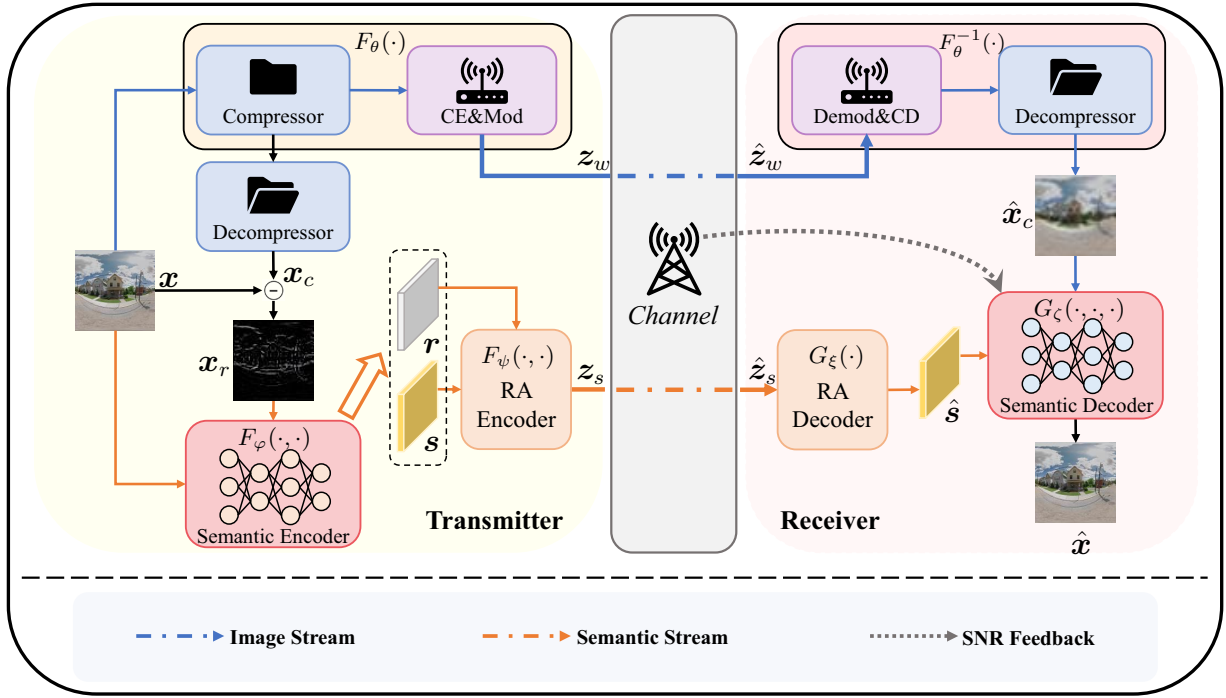
	
	\subsection{General System Model}\label{sec2 system model}
	Consider that a source vector $\boldsymbol{x}\in\mathbb{R}^k$ is transmitted over a wireless channel to the receiver for recovery. In the scenario of wireless image transmission, we have $k=H\times W \times C$, where $H$, $W$, and $C$ are the height, width, and channel number of the source image, respectively. The transmitter encodes and modulates $\boldsymbol{x}$ into a complex-valued vector $\boldsymbol{z}\in\mathbb{C}^{n}$. The \emph{channel bandwidth ratio} is hence defined as CBR$=\frac{n}{k}$, which reflects the system's utilization of channel resources per symbol. Before passing through the channel, $\boldsymbol{z}$ is normalized to satisfy $\frac{{\boldsymbol{z}^{*}}\boldsymbol{z}}{n}\le P$, where $P$ is the maximum transmission power. By $p_{\hat{\bm{z}}|\bm{z}}(\hat{\bm{z}}|\bm{z})$ we denote the transition probability of the wireless channel. Then, after transmission over the noisy channel, the receiver has the corrupted sequence $\boldsymbol{\hat{z}}$ and attempts to reconstruct $\boldsymbol{\hat{x}}$ from $\boldsymbol{\hat{z}}$.

	\subsection{Depiction of ParaSC}
	The proposed ParaSC works under the scenario described in Sec.~\ref{sec2 system model}. As is displayed in Fig.~\ref{fig:framework}, our framework follows a parallel-branch architecture. While the conventional branch obeys an SSCC fashion, the semantic branch is composed of a residual-enhanced semantic encoder, a conditional rate-adaptive module, and an integrated semantic decoder. The image stream $\bm{z}_w$ is transmitted with the semantic stream $\bm{z}_s$ together, which means $\bm{z}=(\bm{z}_w,\bm{z}_s)$.
	
	\subsubsection{Conventional Branch}
	The conventional branch of ParaSC adopts the separation-based coding scheme. The original image $\bm{x}$ is first converted into a binary sequence via a conventional compression method (e.g., JPEG, JPEG2000, BPG), where the source coding ratio is controlled by quality factor $q$. On one hand, this sequence is instantaneously decompressed into the image $\bm{x}_c$ for residual computation. On the other hand, a copy of the binary sequence is input to a conventional channel encoder and is modulated as the image stream $\bm{z}_w=F_{\theta}(\bm{x})\in\mathbb{C}^m$ for transmission\footnote{For convenience, we denote the decoder at the receiver $F^{-1}_{\theta}(\cdot)$.}, where the non-learnable parameter $\theta$ consists of the compression quality factor, channel coding rate, and modulation method. Then, the stream is transmitted through the channel, received as $\hat{\bm{z}}_w\in\mathbb{C}^m$, demodulated and decoded successively to produce the image $\hat{\bm{x}}_c$. Note that the image codec is generally non-differentiable hence it remains invariant during the training process. The encoding and decoding can be expressed as
	\begin{equation}
		\bm{x}\xrightarrow{F_{\theta}(\cdot)}\bm{z}_w,	\quad \hat{\bm{z}}_w\xrightarrow{F^{-1}_{\theta}(\cdot)}\hat{\bm{x}}_c.\notag
	\end{equation}
	\subsubsection{Semantic Branch}
	The semantic branch of ParaSC performs JSCC on the source data and a flexible rate adaptation of semantic information. Having obtained the decompressed image $\bm{x}_c$, the semantic encoder $F_{\varphi}(\cdot, \cdot): \mathbb{R}^k\times\mathbb{R}^k\mapsto\mathbb{R}^{n_s}\times\mathbb{R}^{n_r}$ extracts the key features via the integration of the image residual $\bm{x}_r=\bm{x}-\bm{x}_c$ and a learnable network. The exploited residual, which mainly represents the high-frequency loss during conventional compression, activates the semantics through a simple attention mechanism. The outputs from the network are the raw semantic information $\bm{s}\in\mathbb{R}^{n_s}$ and the residual features $\bm{r}\in\mathbb{R}^{n_r}$, respectively. At the receiver, the semantic decoder takes both the decoded image $\hat{\bm{x}}_c$ and the recovered semantics $\hat{\bm{s}}$ as inputs to restore the image as $\hat{\bm{x}}$. During decoding, the signal-to-noise ratio (SNR) is utilized to determine the policy for parallel-stream integration. We define the decoder mapping as $G_{\zeta}(\cdot, \cdot, \cdot): \mathbb{R}^k\times\mathbb{R}^{n_s}\times\mathbb{R}\mapsto\mathbb{R}^k$. The entire JSCC encoding and decoding process can be written as	
	\begin{align}
	\left(\bm{x},\bm{x}_r\right)&\xrightarrow{F_{\varphi}(\cdot, \cdot)}\left(\bm{s}, \bm{r}\right),\notag\\
	\left(\hat{\bm{x}}_c,\hat{\bm{s}},\mbox{SNR}\right)&\xrightarrow{G_{\zeta}(\cdot, \cdot, \cdot)}\hat{\bm{x}}.\label{eq:decoder}
	\end{align}
	
	Aiming at more efficient semantic transmission, the rate-adaptive encoder (RA Encoder) compresses raw semantic information $\bm{s}$ conditioned with residual $\bm{x}_r$. Specifically, we treat residual features as residual hyperprior, which is a prior on the parameters of the entropy model of the raw semantic information $\bm{s}$. By calculating the conditional entropy of $\bm{s}$ given the residual, we determine the minimum required rate to transmit each semantic element and transform $\bm{s}$ into the semantic stream $\bm{z}_s\in\mathbb{C}^{(n-m)}$. $\bm{z}_s$ is first converted into a complex-value sequence by combining two successive real-value symbols into one complex pair, and then delivered through the channel and recovered by the rate-adaptive decoder (RA Decoder) as $\hat{\bm{s}}$. We define the mapping of RA Encoder $F_{\psi}(\cdot, \cdot):\mathbb{R}^{n_s}\times\mathbb{R}^{n_r}\mapsto\mathbb{C}^{n-m}$ and that of RA Decoder $G_{\xi}(\cdot):\mathbb{C}^{(n-m)}\mapsto\mathbb{R}^{n_s}$, thus formulating the learnable rate adaptation processes as
	\begin{equation}
	\left(\bm{s}, \bm{r}\right)\xrightarrow{F_{\psi}(\cdot, \cdot)}\bm{z}_s, \quad \bm{\hat{z}}_s\xrightarrow{G_{\xi}(\cdot)}\hat{\bm{s}}.\notag
	\end{equation}
	\subsubsection{Wireless Transmission} The semantic stream $\bm{z}_s$ is transmitted with the image stream $\bm{z}_w$ simultaneously through two separate but not necessarily independent channels. Then, we obtain distorted sequences $\boldsymbol{\hat{z}}_w$ and $\boldsymbol{\hat{z}}_s$. The above process can be formulated as
	\begin{align}
		&(\bm{z}_w,\bm{z}_s)\xrightarrow{p_{\hat{\bm{z}}|\bm{z}}} (\boldsymbol{\hat{z}}_w,\boldsymbol{\hat{z}}_s),\notag
	\end{align}
	where $\hat{\bm{z}}=(\hat{\bm{z}}_w,\hat{\bm{z}}_s)$. In this paper, we model the transfer function of the channel as
	\begin{equation}
	\boldsymbol{\hat{z}}=\boldsymbol{h} \odot \boldsymbol{z}+\boldsymbol{n},\label{eq:channel}
	\end{equation}
	where $\boldsymbol{h}$ is the channel gain vector, $\boldsymbol{n}\sim\mathcal{CN}\left(\mathbf{0}, \sigma^2\boldsymbol{I}_n\right)$ denotes the additive white Gaussian noise (AWGN) with average power $\sigma^2$, and $\odot$ denotes the component-wise multiplication.

	\section{JSCC Module for Semantic Branch}\label{Sec3}
	In this section, we detail the design of semantic JSCC in the proposed ParaSC. Specifically, at the transmitter, an attention-biased lightweight module generates a learnable semantic stream with the assistance of the image residual. At the receiver, we first extract the latents of the image stream and then integrate it with the semantic stream. Moreover, an SNR-aware mechanism is adopted to promote the dynamic aggregation of the parallel streams and mitigate the negative channel influence.

	\subsection{The Proposed Semantic Encoder}
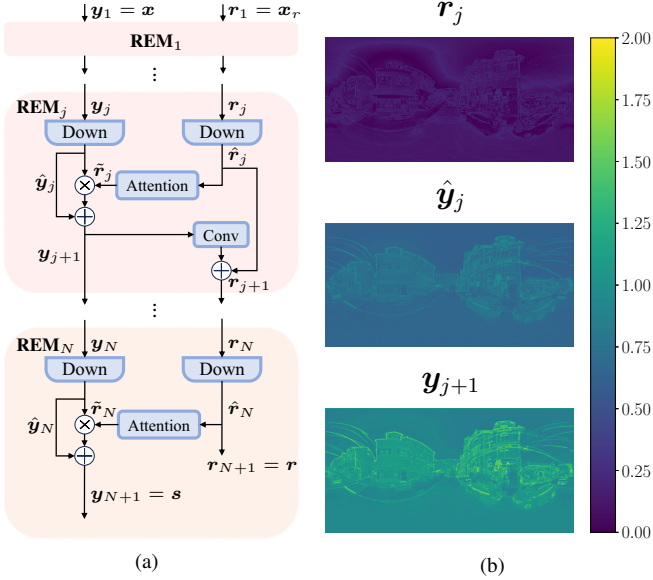
\begin{figure}[tbp]
	\begin{minipage}[tbp]{0.46\linewidth}
		\subfloat[]{
			\centering
			\includegraphics[width=1\textwidth]{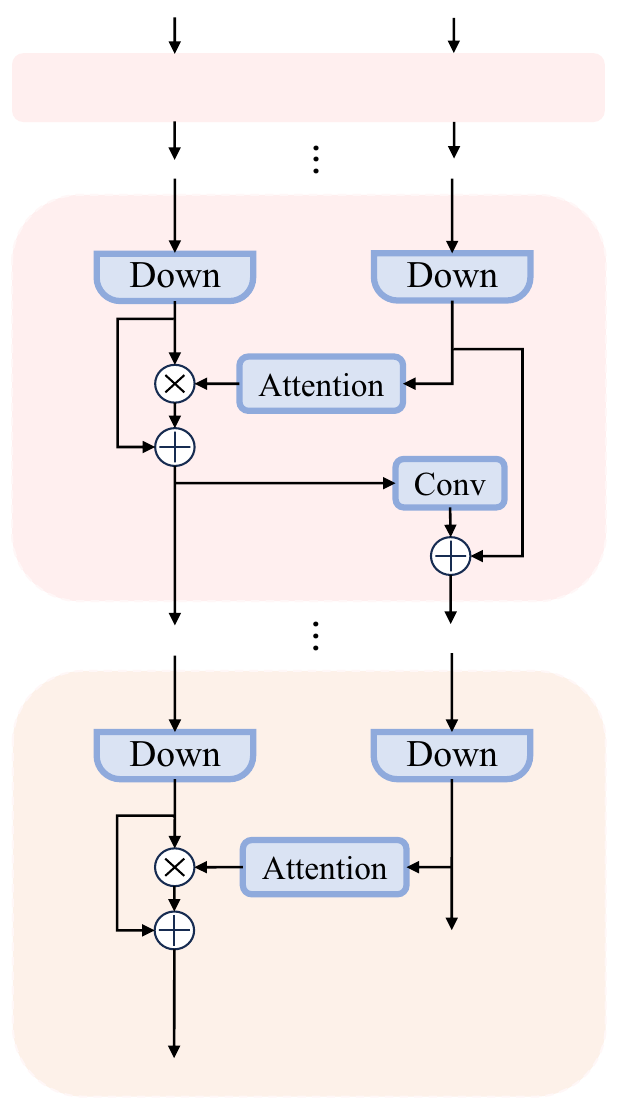}\label{fig:semantic encoder}}
			  \begin{picture}(100,0)
				{\scriptsize
					\put(-66,187){$\textbf{REM}_1$}
					\put(-81,199){$\bm{y}_1=\bm{x}$}
					\put(-29,199){$\bm{r}_1=\bm{x}_r$}
					
					\put(-109,162){$\textbf{REM}_j$}
					\put(-81,163){$\bm{y}_j$}
					\put(-29,163){$\bm{r}_j$}
					\put(-80,138){$\tilde{\bm{r}}_j$}
					\put(-29,144){$\hat{\bm{r}}_j$}
					\put(-102,134){$\hat{\bm{y}}_j$}
					\put(-29,96){$\bm{r}_{j+1}$}
					\put(-101,107){$\bm{y}_{j+1}$}

					\put(-109,72){$\textbf{REM}_N$}
					\put(-81,73){$\bm{y}_N$}
					\put(-29,73){$\bm{r}_N$}
					\put(-81,48){$\tilde{\bm{r}}_N$}
					\put(-29,48){$\hat{\bm{r}}_N$}
					\put(-105,42){$\hat{\bm{y}}_N$}
					\put(-37,27){$\bm{r}_{N+1}=\bm{r}$}
					\put(-81,16){$\bm{y}_{N+1}=\bm{s}$}			
				}
				\end{picture}
	\end{minipage}
	\begin{minipage}[tbp]{0.54\linewidth}
	\subfloat[]{
		\centering
		\begin{annotationimage}{width=0.985\textwidth}{residual_effect_raw.eps}\label{fig:residual effect}
		\draw[coordinate label = {$\bm{r}_j$ at (0.37,0.96)}];
		\draw[coordinate label = {$\hat{\bm{y}}_j$ at (0.37,0.63)}];
		\draw[coordinate label = {$\bm{y}_{j+1}$ at (0.37,0.29)}];
		\end{annotationimage}
		}
		\end{minipage}  
		\caption{(a) The structure of the semantic encoder, where the left pathway processes image and the right pathway processes image residual. (b) Visualization of the residual attention mechanism. Compared to $\hat{\bm{y}}_j$, the high-frequency components are highlighted in $\bm{y}_{j+1}$.}
	\end{figure}

	The residual $\bm{x}_r$ always carries the compression loss arising from the conventional SSCC, which allows the semantic encoder to focus on the features neglected in the image compression. To be specific, in JPEG, these neglected features usually refer to the high-frequency components, which are abandoned by the unequal quantization of the Discrete Cosine Transform (DCT) coefficients. Therefore, the residual provides the encoder with this additional information to enhance the semantic features. To realize this, we treat the image and the residual as two correlated sources for parallel processing, where the encoded residual serves as an attention map acting on the extracted features. The overall architecture of the semantic encoder is illustrated in Fig.~\ref{fig:semantic encoder}, where $\boldsymbol{x}$ and $\boldsymbol{x}_r$ are adopted as initial inputs and passed through $N$ consecutive residual-enhanced modules (REM) to extract the semantic representations iteratively. The iteration steps of $F_{\varphi}(\cdot, \cdot)$ can be expressed as follows
	\begin{align}
		&\bm{y}_1=\bm{x}, \bm{r}_1 = \bm{x}_r,\notag\\
		&\left(\bm{y}_{j+1}, \bm{r}_{j+1}\right) = \mathrm{REM}_{j}(\bm{y}_{j}, \bm{r}_{j}),\quad j=1,2,\cdot\cdot\cdot,N, \notag
	\end{align}
	and the module then outputs semantic and residual features as\footnote{Herein, the cardinality of $j$ denotes the number of successive REMs. Note that we remove the subscript of the output features for simplicity.} $\left(\bm{s}, \bm{r}\right) = \left(\bm{y}_{N+1}, \bm{r}_{N+1}\right)$.

	$\mathrm{REM}_j(\cdot,\cdot)$ is a two-pathway neural network, with one path processing $\bm{y}_{j}$ and the other one processing $\bm{r}_{j}$. Inter-path connections are realized through an attention block $\mathrm{Attention}(\cdot)$ and a convolution layer $\mathrm{Conv(\cdot)}$ with kernel size 1. First, two $\mathrm{Down}(\cdot)$ modules are applied on $\bm{y}_{j}$ and $\bm{r}_{j}$, respectively, where $\mathrm{Down}(\cdot)$ consists of two convolution layers. These layers down-sample on the spatial scale and increase the channels of the features and then output $\hat{\bm{y}}_{j}$ and $\hat{\bm{r}}_{j}$. Second, the attention block $\mathrm{Attention}(\cdot)$ comprising a convolution layer extracts the residual signal of $\boldsymbol{x}$ to produce the attention map $\Tilde{\bm{r}}_{j}$. $\hat{\bm{y}}_{j}$ is then selectively enhanced by $\Tilde{\bm{r}}_{j}$ through an element-wise multiplication and outputs $\bm{y}_{j+1}$. The residual attention enhancement in the $j$-th REM can be expressed as
	\begin{align}
		\bm{y}_{j+1} = \bm{\hat{y}}_j + \bm{\hat{y}}_j \odot \mathrm{Attention}\left(\bm{\hat{r}}_j\right). \notag
	\end{align}
	
	 $\bm{y}_{j+1}$ also participates in the computation of $\bm{r}_{j+1}$ by the $\mathrm{Conv}(\cdot)$ module. Note that $\mathrm{Conv}(\cdot)$ is removed from the last REM to directly output $\bm{r}_{N+1}=\hat{\bm{r}}_N$.
	
	The core of the semantic encoder is $\mathrm{Attention}(\cdot)$, which fuses the residual information into semantic extraction. To verify its effectiveness, we choose one image sample and visualize its latents during the JSCC encoding in Fig. \ref{fig:residual effect}, where $\bm{r}_j$, $\bm{\hat{y}}_j$ and $\bm{y}_{j+1}$ represent the residuals, semantics before $\mathrm{Attention}(\cdot)$, and semantics after $\mathrm{Attention}(\cdot)$, respectively. It is clear that the edges and contours, which are abandoned during the lossy image compression, are precisely activated in $\bm{y}_{j+1}$, with the approximate invariant of low-frequency regions. Therefore, our residual-based encoder shows great potential for image semantic extraction.
	
	\subsection{The Proposed Semantic Decoder}
\begin{figure}[tbp]
		\centering
		\includegraphics[width=0.5\textwidth]{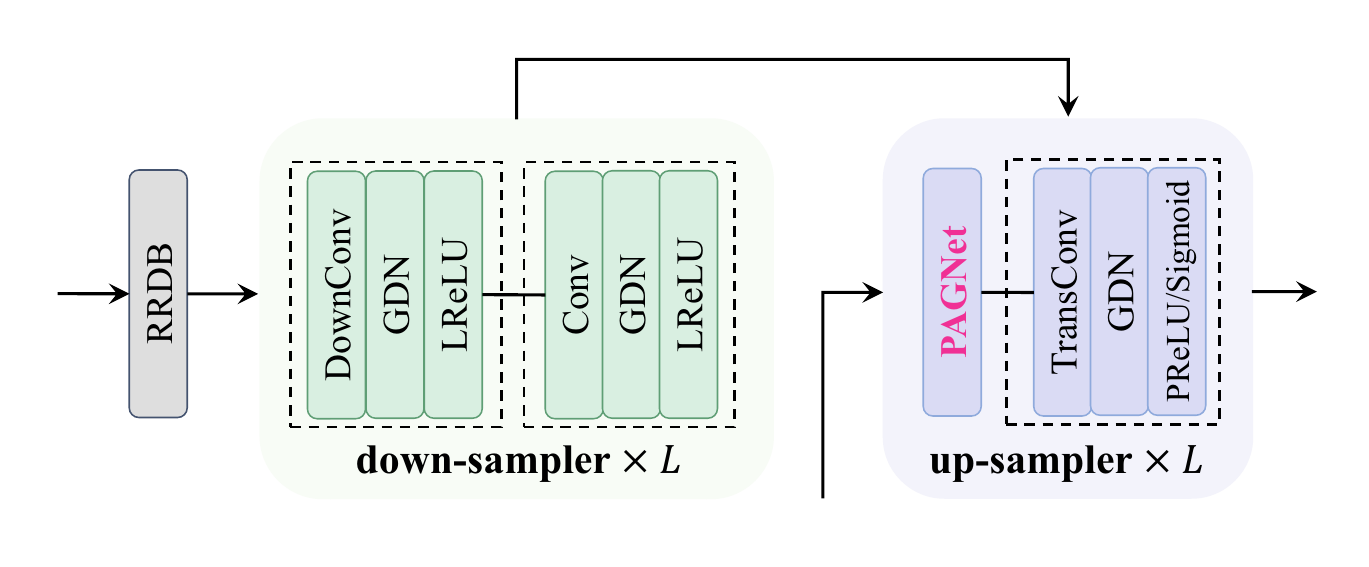}	
		\begin{picture}(100,0)
	{\small
		\put(-75,61){$\hat{\bm{x}}_c$}
		\put(-39,67){$\bm{u}_0$}
		\put(45,110){$\bm{u}_1,\bm{u}_2,\cdots,\bm{u}_L$}
		\put(72,17){$\hat{\bm{s}},\mbox{SNR}$}
		\put(174,61){$\hat{\bm{x}}$}
	}
\end{picture}
		\caption{The architecture of the semantic decoder, which includes a novel PAGNet in the up-sampling block. The "DownConv" and "TransConv" denote the convolution layer and transpose convolution layer, respectively.}
		\label{fig:semantic decoder}
	\end{figure}
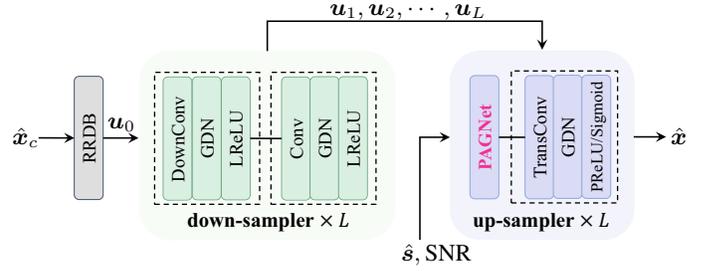

	Given received semantic features $\hat{\bm{s}}$ and possibly corrupted decoded image $\hat{\bm{x}}_c$, the semantic decoder recovers the image $\boldsymbol{\hat{x}}$ to retain visual quality and task-beneficial semantic knowledge. Inspired by UNet~\cite{unet} which can fuse the features through skip connections, we propose a fusion decoder structure, which is shown in Fig.~\ref{fig:semantic decoder}. The semantic decoding includes two steps: \emph{Latent extraction} and \emph{Semantic-integrated Reconstrcution}. First, latent vectors of the conventional compressed image $\hat{\bm{x}}_c$ are extracted to capture the multi-level representations of the image stream. Second, several up-sampling blocks based on transpose convolution reconstruct the image with deep semantic feature $\bm{\hat{s}}$, during which parallel aggregation networks (PAGNet) realize a dynamic combination of two streams to enhance the robustness of the framework.
	\subsubsection{Latent Extraction}
	The process of latent extraction can be expressed as
	\begin{align}
		\bm{u}_0 &= G_{\gamma}(\boldsymbol{\hat{x}_c}), \notag\\
		\bm{u}_{\ell} &= 
			G_{\eta}\left(\bm{u}_{\ell-1}\right), &\ell = 1,2,\cdot\cdot\cdot,L, \label{latent} 
	\end{align}
	where $G_{\gamma}(\cdot)$ is the residual-in-residual dense block (RRDB)~\cite{rrdb}, $G_{\eta}(\cdot)$ denotes a down-sampling block consisting of two convolution layers, two Generalized Divisive Normalization~\cite{gdn} (GDN) layers, and two LeakyReLU functions. RRDB first pre-processes $\boldsymbol{\hat{x}}_c$ to obtain the feature $\bm{u}_0$, followed by $L$ successive down-samplers to iteratively reduce the spatial scale. After each convolution layer, a GDN layer is applied to normalize the output before sending it to the activation function. The downscaling ratio of each $G_{\eta}$ is set to 2.

	\subsubsection{Semantic-integrated Reconstruction}
	 We utilize the extracted latents in \eqref{latent} and the transmitted semantics $\bm{\hat{s}}$ to fuse and up-sample the latents to generate the image. For $\ell \in \{1, 2,\cdot\cdot\cdot,L\}$, the up-sampling step can be described as
    \begin{align}
		\bm{v}_{\ell} &= G_{\nu}\left(\bm{u}_{L+1-\ell}, \bm{v}_{\ell-1}, \mbox{SNR}\right), \notag
    \end{align}
    where $\bm{v}_0=\bm{\hat{s}}$, $\bm{v}_L=\bm{\hat{x}}$, and $G_{\nu}$ encapsulates all the layers of an up-sampling block, including a parallel aggregation network (PAGNet), a transpose convolution layer, a GDN layer and, an activation function which is Parametric ReLU for $\ell \in \{1,2,\cdot\cdot\cdot,L-1\}$ and Sigmoid for $\ell=L$. We have $\zeta=(\gamma,\eta,\nu)$, where $\zeta$ is defined in Eq.~\eqref{eq:decoder} to represent all the parameters of semantic decoder.
    \begin{figure}[tbp]
	\centering
	\includegraphics[width=0.35\textwidth]{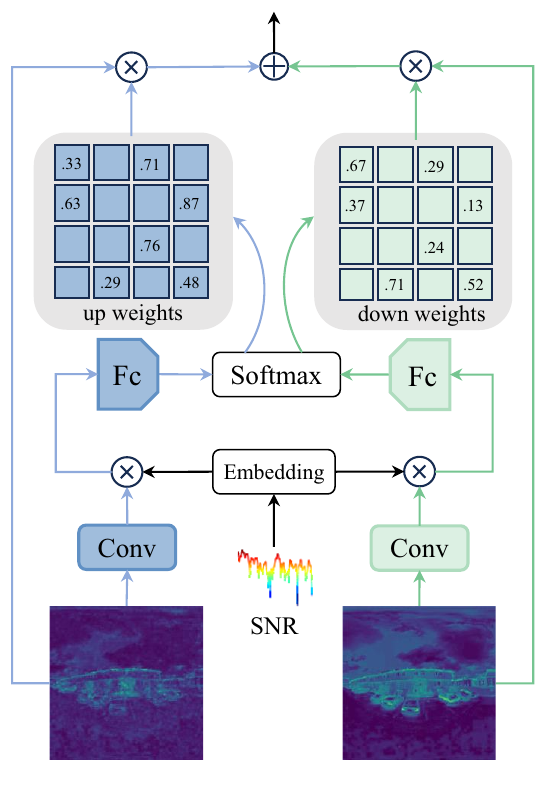}
	\begin{picture}(80,0)
		{\normalsize
			\put(-18,18){$\bm{v}_{\ell-1}$}
			\put(74,18){$\bm{u}_{L+1-\ell}$}
		}
	\end{picture}
	\caption{Block diagram of PAGNet in the $\ell$-th up-sampler, where the left pathway deals with the up-sampled latents (starting with $\bm{v}_0=\bm{\hat{s}}$) and the right pathway processes the down-sampled latents (extracted from decoded image $\hat{\bm{x}}_c$).}
	\label{pagnet}
\end{figure}

    The motivation behind PAGNet is that generally, there exists an SNR threshold where the received image stream becomes too erroneous for the channel decoder to recover, leading to a corrupted image $\hat{\bm{x}}_c$ which brings detrimental effects to the reconstruction; on the contrary, the semantic features extracted by neural networks are more resilient to channel noise. When the channel is quite noisy, it is better to let the semantic stream dominate the reconstruction process. To achieve this, PAGNet utilizes the SNR feedback to assign different aggregation weights to the image and semantic streams based on different channel states. The weights are pixel-level instead of instance-level, which is more flexible in preserving the clear regions and inhibiting the noisy regions of the compressed image.
    
     Fig.~\ref{pagnet} depicts the structure of PAGNet. First, each latent goes through a convolution layer to analyze the noisy patterns. The received SNR is transformed by an embedding layer to an SNR vector, followed by an element-wise product with the latent and a fully-connected (FC) layer to get the raw weights. The Softmax function normalizes these weights in a per-pixel manner, generating the final weights to aggregate them. Since the latents have $L$ different scales, every up-sampling block in the semantic decoder is assigned with a specialized PAGNet, which does not share its parameters with others.

\section{Residual-conditioned Semantic Rate Adaptation}\label{Sec4}
As is mentioned above, ParaSC realizes the integration of conventional SSCC and DL-driven JSCC by simultaneous transmission of the image stream and semantic stream. So far, the transmission rate is only controllable in the image stream by configuring the parameters of the conventional SSCC. The semantic rate is fixed once the JSCC parameters are determined, which results in an inevitable performance loss when the rate of the other stream varies. To further improve the efficiency of JSCC transmission, in this section, we introduce a semantic rate adaptation mechanism according to diverse coding conditions of the SSCC. 

Since most of the information of the original image has been transmitted by the image stream from the SSCC, we only need to extract the semantic features from what was neglected by SSCC, namely, the residual. This residual information is characterized by the residual hyperprior $\bm{r}$, an indicator to help further reduce the semantic redundancy. Therefore, we control the semantic rate by adjusting the output length of the extracted semantic information $\bm{s}$ according to the entropy model established by residual hyperprior. Specifically, we estimate the likelihood of $\bm{s}$ to calculate its conditional entropy as the minimum transmission rate. The estimation is performed based on~\cite{ntscc}, but the major difference is that our work uses the residual as the condition to generate the hyperprior before the rate allocation stage, which enables the rate adaptation of the semantic stream conditioned on the image stream and also spares the framework from introducing an extra prior. 

In this section, we first analyze the rate-distortion (RD) tradeoff from the perspective of conditional variational auto-encoder (CVAE) and then propose our rate-adaptive codec.
\begin{figure*}[htbp]
	\centering
	\includegraphics[width=0.85\textwidth]{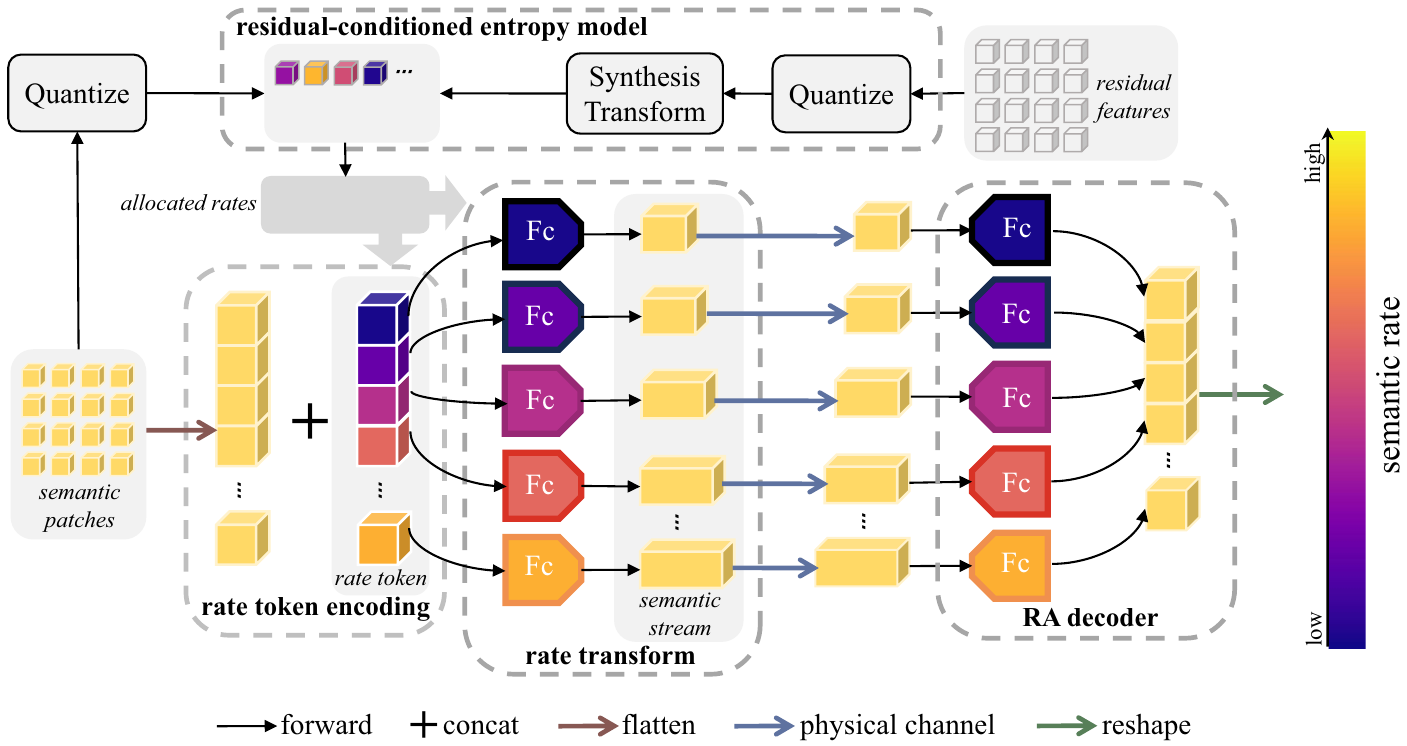}
	\begin{picture}(0,0)
		{\scriptsize
			\put(-359,195){$-\log p (\tilde{\bm{s}},\tilde{\bm{r}}|\bm{x}_r)$}
			\put(-360,168){$\bar{\alpha}_1,\bar{\alpha}_2,\cdots,\bar{\alpha}_{k_s}$}
		\large
			\put(-388,207){$\tilde{\bm{s}}$}
			\put(-292,208){$\tilde{\bm{\mu}},\tilde{\bm{\sigma}}$}
			\put(-211,208){$\tilde{\bm{r}}$}
			\put(-40,106){$\hat{\bm{s}}$}		
		}
	\end{picture}
	\caption{The conditional rate adaptation module in ParaSC, including a conditional entropy model, an RA encoder and an RA decoder. The RA encoder performs rate token encoding and rate transform successively The rate allocation is conditioned on the residual $\bm{x}_r$, which characterizes the compression loss of the conventional image stream.}
	\label{fig:rate}
\end{figure*}
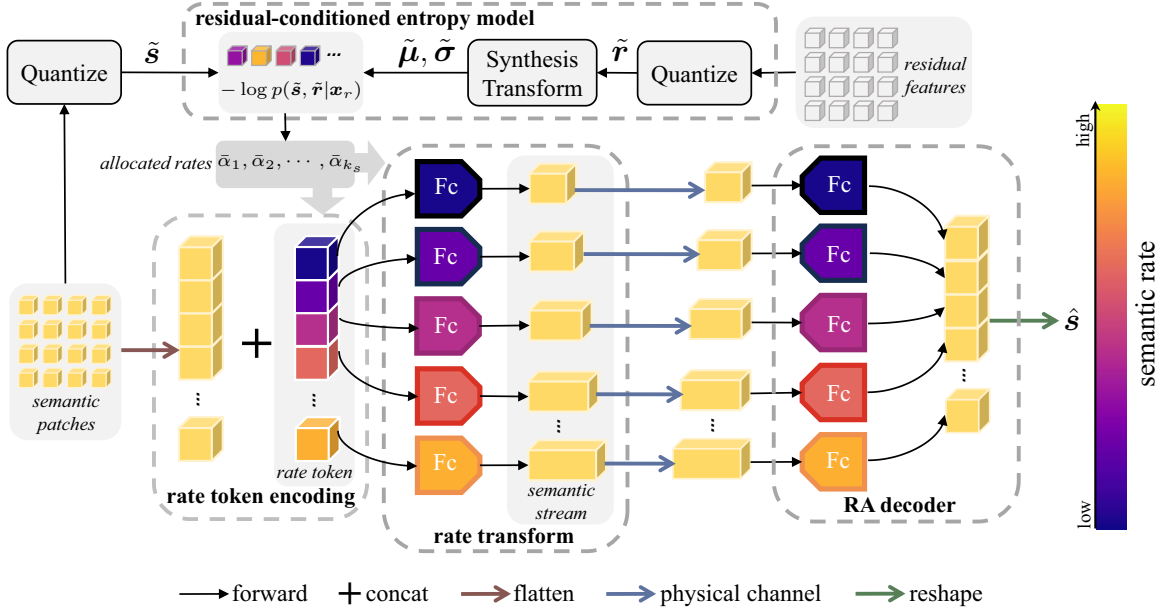
\subsection{Theoretical Analysis}

By modifying the evidence lower bound (ELBO) \cite{Blei2022}, we start from the KL divergence between the estimated probability distribution 
$q_{\varphi,\psi}(\hat{\bm{z}}_s, \bm{\tilde{r}}|\bm{x},\bm{x}_r)$ and the true latent distribution $p_{\varphi,\psi}(\hat{\bm{z}}_s, \bm{\tilde{r}}|\bm{x},\bm{x}_r)$ by conditioning the residual features derived from the conventional encoding scheme\footnote{For brevity, we omit the parameter in the subscript of the probability distribution when possible.}
\begin{align}
	&D_{\mathrm{KL}}\left[q_{\varphi,\psi}(\hat{\bm{z}}_s, \bm{\tilde{r}}|\bm{x},\bm{x}_r)||p_{\varphi,\psi}(\hat{\bm{z}}_s, \bm{\tilde{r}}|\bm{x},\bm{x}_r)\right]\notag\\
	=&\mathbb{E}_{q(\hat{\bm{z}}_s, \tilde{\bm{r}}|\bm{x},\bm{x}_r)}\left[\log\frac{q(\hat{\bm{z}}_s,\tilde{\bm{r}}|\bm{x},\bm{x}_r)}{p(\hat{\bm{z}}_s,\tilde{\bm{r}}|\bm{x},\bm{x}_r)}\right]\notag\\
	=&\mathbb{E}_{q(\hat{\bm{z}}_s, \tilde{\bm{r}}|\bm{x},\bm{x}_r)}[\underbrace{\log q(\hat{\bm{z}}_s,\tilde{\bm{r}}|\bm{x},\bm{x}_r)}_{T_1}\underbrace{-\log p\left(\bm{x}|\hat{\bm{z}}_s,\tilde{\bm{r}},\bm{x}_r\right)}_{T_2}\notag\\&\hspace{2cm}\underbrace{-\log p(\hat{\bm{z}}_s,\tilde{\bm{r}}|\bm{x}_r)}_{T_3}\underbrace{+\log p(\bm{x}|\bm{x}_r)}_{T_4}],~\label{kl}
\end{align}
where $\tilde{\bm{r}}$ denotes the quantized residual hyperprior. During the model training, we replace the hard quantization with a relaxed form to allow stochastic gradient descent~\cite{ballé2017endtoend}, which is achieved by adding a uniform noise vector $\bm{o}_r\sim \mathcal{U}(-\frac{1}{2},\frac{1}{2})$ to $\bm{r}$. Denote $\round{\cdot}$ as the uniform scalar quantization which quantizes each element to the nearest interger. We have
\begin{equation}
  \tilde{\bm{r}}=\left\{
  \begin{aligned}
  \bm{r}+\bm{o}_r,\quad &\text{in model training},\\
  \round{\bm{r}},\quad &\text{in model testing}.
  \end{aligned}
  \right.\label{eq:quantization}
\end{equation}

 In~\eqref{kl}, the term $\mathbb{E}_q[T_1]$ corresponds to the joint posterior by performing semantic encoding and rate-adaptive encoding\footnote{In the subsequent paragraphs, $\mathbb{E}_{q}[\cdot]$ refers to $\mathbb{E}_{q(\hat{\bm{z}}_s, \tilde{\bm{r}}|\bm{x},\bm{x}_r)}[\cdot]$ unless otherwise specified.}, then adding uniform noise and channel effects, we have
\begin{align}
	q(\hat{\bm{z}}_s, \tilde{\bm{r}}|\bm{x},\bm{x}_r)&= \prod_{i=1}^{n-m} \mathcal{CN}(\hat{z}_{s_i}|z_{s_i},\sigma^2)\notag\\
	&\cdot \prod_{j=1}^{J}\mathcal{U}(\tilde{r}_j|r_j-\frac{1}{2},r_j+\frac{1}{2})\notag\\
	\mbox{with} \ (\bm{s},\bm{r})&=F_{\varphi}(\bm{x},\bm{x}_r), \bm{z}_s=F_{\psi}(\bm{s},\bm{r}),\notag
\end{align}
where $\hat{z}_{s_i}$ and $z_{s_i}$ denotes the $i$-th element of $\hat{\bm{z}}_{s}$ and $\bm{z}_{s}$, respectively. Following~\cite{ntscc}, the uniform distribution has a constant support width and thus $\mathbb{E}_q[T_1]$ is constant. Hence $\mathbb{E}_q[T_1]$ can be dropped from~\eqref{kl}.

Using Jensen inequality, the second term $\mathbb{E}_q[T_2]$ has the upper bound
\begin{align*}
	-\log p(\bm{x}|\hat{\bm{z}}_s,\tilde{\bm{r}},\bm{x}_r)&=-\log \mathbb{E}_{p(\bm{s}|\hat{\bm{z}}_s,\tilde{\bm{r}})}\left[p(\bm{x}|\bm{s},\bm{x}_r)\right]\\\notag
	&\le -\mathbb{E}_{p(\bm{s}|\hat{\bm{z}}_s,\tilde{\bm{r}})}\left[\log p(\bm{x}|\bm{s},\bm{x}_r)\right]\\\notag
	&=-\mathbb{E}_{p(\bm{s}|\hat{\bm{z}}_s,\tilde{\bm{r}})}\left[\log p(\bm{x}|\bm{s},\bm{x}_c)\right]\\\notag
	&=-\mathbb{E}_{p(\bm{s}|\hat{\bm{z}}_s,\tilde{\bm{r}})}\left[\log p(\bm{x}|\bm{s},\hat{\bm{x}}_c)\right]\\\notag
	p(\bm{x}|\bm{s},\hat{\bm{x}}_c)&=\mathcal{N}(\bm{x}|\hat{\bm{x}},2\lambda^{-1}\mathbf{I}_k)\\\notag
	&\mbox{with} \ \hat{\bm{x}} =G_{\zeta}\left(\hat{\bm{x}}_c,\bm{s},\mbox{SNR}\right),\notag
\end{align*}
where $\lambda$ is the weighted coefficient on the distortion term. The condition $\bm{x}_r$ retains the high-frequency components of a specific original image $\bm{x}$, which are distinguishable from those of other images. Therefore, $\bm{x}_r$ determines a unique $\bm{x}$ although such mapping is non-injective. Then, the equation $\bm{x}_c=\bm{x}-\bm{x}_r$ renders $\bm{x}_c$ a known condition. $\hat{\bm{x}}_c$ is the decompressed image after transmission, where the channel state is fixed during the training stage. Hence $\bm{\hat{x}}_c$ can also be regarded as the condition. $\mathbb{E}_q[T_2]$ stands for the mean square error (MSE) between the reconstructed image $\hat{\bm{x}}$ and the original image $\bm{x}$.

The thrid term $\mathbb{E}_q[T_3]$ is the cross-entropy encoding $\hat{\bm{z}}_s$ and $\tilde{\bm{r}}$ given the residual $\bm{x}_r$ according to the conditional entropy model. The fourth term $\mathbb{E}_q[T_4]$ reflects the distortion which merely arises from the conventional image codec, which cannot be optimized because the conventional branch is assumed to be not trainable. Let $\mathbb{E}_q[T_1+T_4]=C$, ~\eqref{kl} can be rewritten as the form of rate-distortion loss
\begin{align}
	&D_{\mathrm{KL}}\left[q_{\varphi,\psi}(\hat{\bm{z}}_s,\tilde{\bm{r}}|\bm{x},\bm{x}_r)||p_{\varphi,\psi}(\hat{\bm{z}}_s,\tilde{\bm{r}}|\bm{x},\bm{x}_r)\right]\notag\\
	\le&\mathbb{E}_{q(\hat{\bm{z}}_s, \tilde{\bm{r}}|\bm{x},\bm{x}_r)}[-\mathbb{E}_{p(\bm{s}|\hat{\bm{z}}_s,\tilde{\bm{r}})}\left[\log p(\bm{x}|\bm{s},\hat{\bm{x}}_c)\right]\notag\\
	&\hspace{3cm}-\log p(\hat{\bm{z}}_s,\tilde{\bm{r}}|\bm{x}_r)]+C\notag\\
	=&\mathbb{E}_{q(\hat{\bm{z}}_s, \tilde{\bm{r}}|\bm{x},\bm{x}_r)}[\underbrace{d(\bm{x},\bm{\hat{x}})}_{{\fontsize{7pt}{8pt}\selectfont \mbox{ distortion}}}\underbrace{-\log p(\hat{\bm{z}}_s,\tilde{\bm{r}}|\bm{x}_r)}_{{\fontsize{7pt}{8pt}\selectfont \mbox{semantic rate}}}]+C,\label{RD1}
\end{align}
where $d(\cdot,\cdot)$ represents MSE (i.e., $d(\bm{x},\hat{\bm{x}})={\Vert \bm{x}-\hat{\bm{x}}\Vert}_2^2$) here.

\subsection{Design of the Conditional Rate Adaptation Module}
Based on $\mathbb{E}_q[T_3]$ in~\eqref{kl}, we design the variable-rate transmission of the semantic stream by estimating the conditional entropy distribution of the semantic features. This entropy model-based rate-adaption mechanism is first presented in \cite{Li2021}, where a Laplacian distribution convolution with a uniform distribution on $(-\frac{1}{2},\frac{1}{2})$ to approximate the conditional likelihood. Following~\cite{ballé2018variational}, we utilize Gaussian distribution to model the conditional semantic and residual features by applying a synthesis transform $h_s$ to $\tilde{\bm{r}}$:
\begin{align*}
	p (\tilde{\bm{s}},\tilde{\bm{r}}|\bm{x}_r)&=\prod_{i=1}^{k_s} \left(p_{\mathcal{N}(\tilde{\mu}_i,{\tilde{\sigma}_i}^2)} * p_{\mathcal{U}(-\frac{1}{2},\frac{1}{2})}\right)(\tilde{s}_i),\\\notag &\mbox{with}\quad(\tilde{\bm{\mu}},\tilde{\bm{\sigma}})=h_s(\tilde{\bm{r}}),
\end{align*}
where $k_s$ denotes the number of patches in $\bm{s}$, and $*$ denotes the convolution. $h_s$ is parameterized by a learnable network with a Convolution-LeakyReLU-Convolution architecture, effectively summarizing the mean value $\tilde{\bm{\mu}}$ and standard deviation value $\tilde{\bm{\sigma}}$ of the conditional semantics. Identical to the residual hyperprior quantization in \eqref{eq:quantization}, we quantize the semantic patches $\bm{s}$ during training by adding a uniform noise $\bm{o}_s \sim \mathcal{U}(-\frac{1}{2},\frac{1}{2})$ and during testing by quantizing them to integers.

Note that the semantic rate term in~\eqref{RD1} is $-\log p(\hat{\bm{z}}_s,\tilde{\bm{r}}|\bm{x}_r)$, where $\hat{\bm{z}}_s$ originates from $\bm{s}$ by rate-adaptive encoding $F_\psi$ and channel transmission $p(\hat{\bm{z}}_s\vert\bm{z}_s)$. While $p(\hat{\bm{z}}_s\vert\bm{z}_s)$ transforms $\bm{z}_s$ through a physical channel with an invariant distribution of the channel parameters during training, $F_\psi$ transforms $\bm{s}$ to $\bm{z}_s$ satisfying the minimum required code lengths determined by the entropy model. The relationship between $p(\bm{z}_s,\tilde{\bm{r}}|\bm{x}_r)$ and $p(\bm{s},\tilde{\bm{r}}|\bm{x}_r)$ is further given by:
\begin{align}
	p(\bm{z}_s,\tilde{\bm{r}}|\bm{x}_r)&\overset{(a)}{=}p(\tilde{\bm{r}}|\bm{x}_r)p(\bm{z}_s|\tilde{\bm{r}},\bm{x}_r)\notag\\
	&\overset{(b)}{=}p(\tilde{\bm{r}}|\bm{x}_r)\int_{\bm{s}}p(\bm{z}_s|\bm{s},\bm{x}_r,\tilde{\bm{r}})p(\bm{s}|\bm{x}_r,\tilde{\bm{r}})d\bm{s}\notag\\
	&\overset{(c)}{=}p(\tilde{\bm{r}}|\bm{x}_r)\int_{\bm{s}}p(\bm{z}_s|\bm{s},\tilde{\bm{r}})p(\bm{s}|\bm{x}_r,\tilde{\bm{r}})d\bm{s}\notag\\
	&\overset{(d)}{=}\int_{\bm{s}}p(\bm{z}_s|\bm{s},\tilde{\bm{r}})p(\bm{s},\tilde{\bm{r}}|\bm{x}_r)d\bm{s}\label{eq:rate transform}
\end{align}
where $p(\bm{z}_s|\bm{s},\tilde{\bm{r}})$ characterizes the transition probability of $F_\psi$. The change from (b) to (c) in \eqref{eq:rate transform} is because once $\bm{s}$ and $\tilde{\bm{r}}$ are given, the distribution of $\bm{z}_s$ will not be correlated with $\bm{x}_r$. As a result, we can observe that optimizing $p(\bm{s},\tilde{\bm{r}}|\bm{x}_r)$ leads to optimizing $p(\hat{\bm{z}}_s,\tilde{\bm{r}}|\bm{x}_r)$ through $F_\psi$ and the channel. During training, we use the soft quantized version $\tilde{\bm{s}}$ to represent the semantic rate term as $-\log p(\tilde{\bm{s}},\tilde{\bm{r}}|\bm{x}_r)$. Hence, by calculating the expectation with respect to $\bm{x}$, the rate-distortion loss is given by
\begin{align}
&\mathcal{L}_{\mathrm{RD}}(\varphi,\psi,\xi,\zeta\vert\theta)\notag\\
=&\mathbb{E}_{p(\bm{x})}\mathbb{E}_{q(\hat{\bm{z}}_s, \tilde{\bm{r}}|\bm{x},\bm{x}_r)}[d(\bm{x},\bm{\hat{x}})-\lambda_1 \log p(\tilde{\bm{s}},\tilde{\bm{r}}|\bm{x}_r)].\label{RD2}
\end{align}
Here, without loss of generality, we change the trade-off coefficient $\lambda$ to $\lambda_1$, which becomes the weight of the semantic rate term. The overall loss is conditioned on the traditional coding scheme $\theta$. The lowest rate of each patch $s_i$ given $\tilde{\bm{r}}$ can be approximated as
\begin{align}
	\alpha_{i}=-\rho\log p(\tilde{s}_i,\tilde{\bm{r}}|\bm{x}_r)\notag \text{ for } i=1,2,\cdots,k_s,
\end{align}
where $\rho$ is a scaling factor. By predefining a rate set $\mathcal{W}=\{w_1,w_2,\cdots,w_M\}$, which incorporates $M$ different integers as the quantized levels, we can obtain the ceiling semantic rate of $s_i$ as
\begin{equation}
	\bar{\alpha}_i = \underset{w_j\ge \alpha_i}{\arg\min}\Vert \alpha_i-w_j \Vert_2,\quad j=1,2,\cdots,M.\label{eq:discrete rate}
\end{equation}
The total bandwidth overhead can be computed as
\[
\mathcal{R}=\frac{1}{k}\left(\sum_{i=1}^{k_s}\bar{\alpha}_i+m\right),
\]
where $m$ is the length of the image stream and thus the corresponding CBR can be determined. 

Fig.~\ref{fig:rate} depicts the architecture of rate adaptation module. First, the semantic features $\bm{s}$ are flattened on the spatial dimensions. Similar to positional encoding, we add corresponding rate tokens to each flatten semantic patch. Then, the residual hyperprior $\bm{r}$ goes through quantization to obtain $\tilde{\bm{r}}$, which is then summarized by a synthesis transform $h_s$ to get the mean and standard deviation value of the conditional distribution $\tilde{\bm{s}}|{\bm{x}_r}$ for a given $\tilde{\bm{r}}$. By applying the conditional Gaussian model, we estimate the log likelihood $-\log p (\tilde{\bm{s}},\tilde{\bm{r}}|\bm{x}_r)$ to determine the conditional entropy of $\bm{s}$, which is further quantized by~\eqref{eq:discrete rate} as the output dimensions of the group of $M$ FC layers to adjust each semantic patch's length. After passing the channel, each patch of the received semantic stream is inversely mapped to its original embedding dimension by another group of FC layers, then reshaped to obtain $\hat{\bm{s}}$. This mechanism hence drives the semantic rate adaption in terms of source distributions and residual conditions.

\section{Simulation Results}\label{Sec5}
In this section, we verify the superior performance of our proposed ParaSC via extensive experiments on the image reconstruction task in comparison with the existing single-stream methods. Results show that ParaSC outperforms these methods in a large range of SNR and CBR for both AWGN and Rayleigh fading channels, especially in perceptual quality metrics. Moreover, ablation studies demonstrate the trade-off between the rates of image stream and semantic stream. Finally, a complexity analysis and visualizations are presented.
\subsection{Experimental Setup}
In this section, we present the experimental details, including the datasets, model details, and training details of ParaSC, and baseline schemes.
\subsubsection{Datasets}
To comprehensively evaluate the performance of our proposed ParaSC as well as baseline schemes, we consider several datasets with different visual representations in our experiments. First, we choose a $360^{\circ}$ field-of-view (FOV) panoramic image dataset CVRG-Pano~\cite{cvrg-pano}, which consists of 524 images for training and 76 images for testing, with a resolution of 1664×832. Second, we choose a standard FOV image dataset Kodak24~\cite{kodak}, which includes 24 RGB images with a resolution of 768×512. Before testing the DL-based methods on Kodak24, they will be trained on the training dataset of DIV2K~\cite{div2k} containing 800 images with a resolution of 2K.

\subsubsection{Model Details}
For the image stream, we select JPEG as the source coding method and adopt a $6144$-bit LDPC with code rate $0.75$ from the 5G standard and QPSK modulation. Maximum likelihood based demodulation and belief propagation based decoding are used. Note that other advanced conventional compression algorithms like JPEG2000 and BPG can also be applied in our framework. For the semantic encoder, the number of REM is set to $N=4$, while the kernel size of all convolution layers is $3\times3$ except for $\mathrm{Conv}$ which is $1\times1$. For the semantic decoder, the numbers of up-sampling and down-sampling blocks are set to $L=4$, while the dimension of the SNR embedding is 32. In the rate-adaptive codec, the FC layers have 32 available equidistant output dimensions ranging from 4 to 128 and we set $\rho=0.2$. The channel input power is normalized to 1. Moreover, we simulate the AWGN channel by setting $\bm{h}$ in~\eqref{eq:channel} to an all-one vector, and the Rayleigh channel by setting each element of $\bm{h}$ as $\mathcal{CN}(0,1)$, with the fading block length matching the code block length.

\subsubsection{Training Details}
According to~\eqref{RD2}, merely regularizing the MSE cannot capture the correlation between the pixels and fails to take the global information of the image into account. Following~\cite{perceptualloss}, we adopt the perceptual loss, which is more beneficial to human perception
\begin{equation}
	\mathcal{L}_{\mathrm{perc}}(\varphi,\psi,\xi,\zeta\vert\theta) = \mathbb{E}_{p(\bm{x})}\left[\frac{1}{k}\Vert \mathrm{\textbf{VGG}}(\boldsymbol{x}) - \mathrm{\textbf{VGG}}(\boldsymbol{\hat{x}}) \Vert_2^{2}\right]\notag,
\end{equation}
where $\mathrm{\textbf{VGG}}(\cdot)$ represents the embedding output of the VGG16~\cite{vgg16} network, which is pre-trained on ImageNet~\cite{imagenet}. Hence, the overall loss function is defined as
\begin{equation}
	\mathcal{L}(\varphi,\psi,\xi,\zeta\vert\theta) = \mathcal{L}_{\mathrm{RD}} + \lambda_{2}\mathcal{L}_{\mathrm{perc}},\label{loss}
\end{equation}
where $\lambda_{2}$ is the weight coefficient of $\mathcal{L}_{\mathrm{perc}}$. Our framework is optimized according to~\eqref{loss} where we set $\lambda_1=\lambda_2=0.1$. We use the Adam optimizer~\cite{Adam} with $\beta_{1}=0.9$ and $\beta_{2}=0.999$. The employed data augmentations include random horizontal flip, random vertical flip, and random crop to $256\times256$.\\

The training of our framework is divided into three stages. In the first stage, we remove the rate-adaptive codec from the framework and train the remaining modules. In the second stage, we freeze the weights trained in the first stage and insert the rate-adaptive codec to train it alone. In the final stage, we finetune all the trainable modules of the framework, where the learning rate is decayed by poly strategy~\cite{liu2015parsenet} with power 0.9 to facilitate convergence.

	\begin{figure*}[htbp]
	\centering
	\begin{minipage}[tbp]{0.3\linewidth}
		\centering
		\subfloat[]{
			\label{PSNR_SNR_AWGN}
			\centering
			\includegraphics[width=1\textwidth]{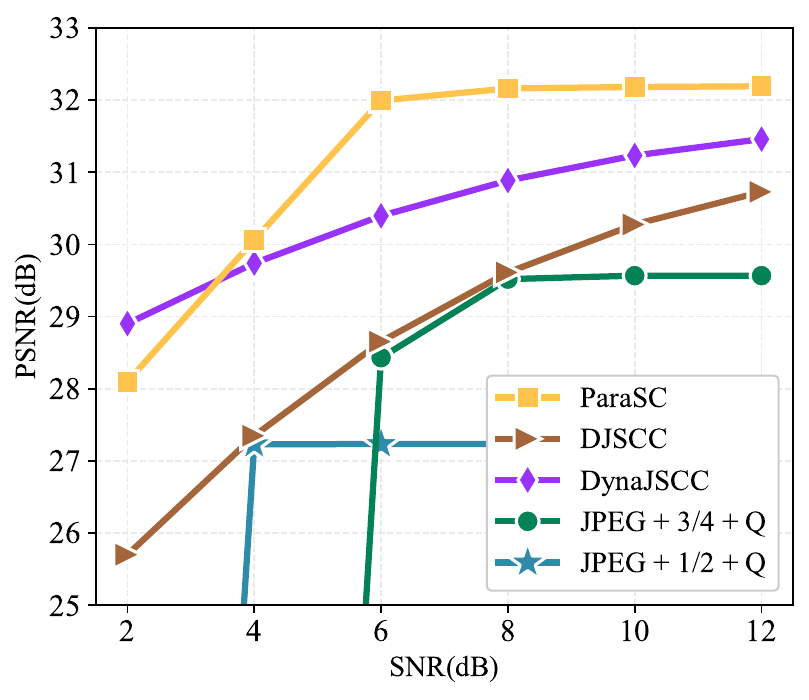}}
	\end{minipage}
	\begin{minipage}[tbp]{0.3\linewidth}
		\centering
		\subfloat[]{
			\label{SSIM_SNR_AWGN}
			\centering
			\includegraphics[width=1\textwidth]{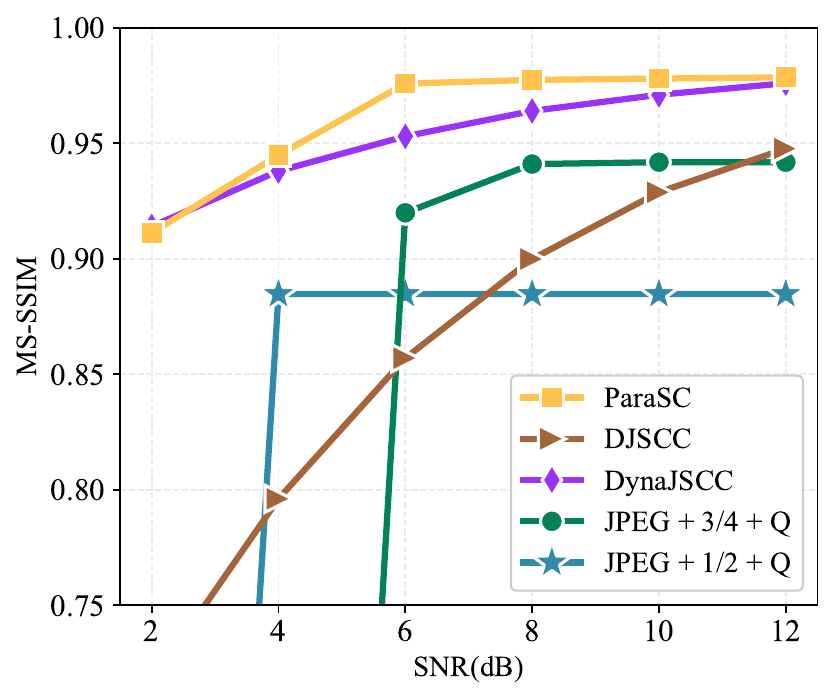}
		}
	\end{minipage}
	\begin{minipage}[tbp]{0.3\linewidth}
		\centering
		\subfloat[]{
			\label{LPIPS_SNR_AWGN}
			\centering
			\includegraphics[width=1\textwidth]{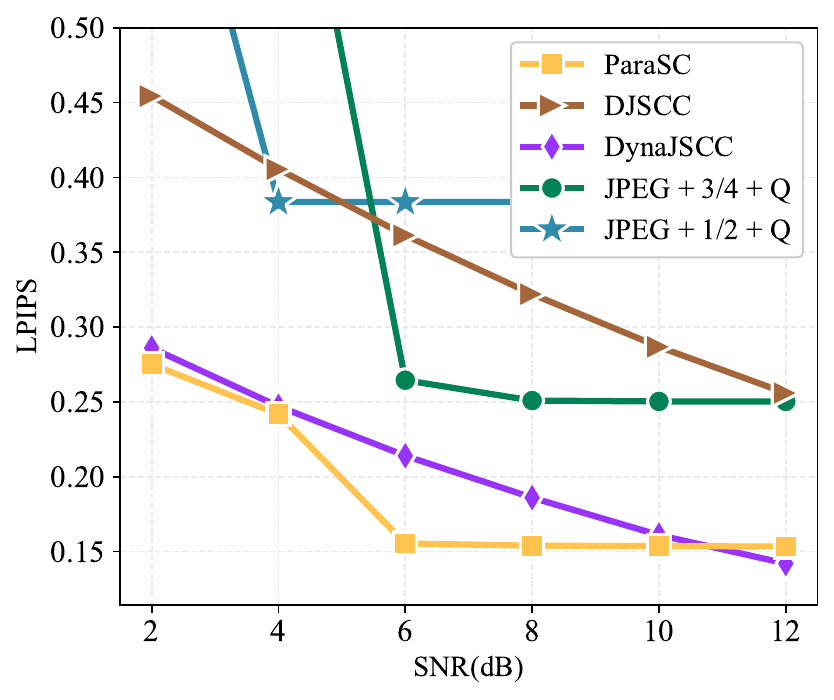}}
	\end{minipage}
	\caption{Comparisons of reconstruction results on CVRG-Pano dataset under AWGN channel in terms of ParaSC, and other baselines with CBR=$0.1$; (a) PSNR against SNR; (b) MS-SSIM against SNR; (c) LPIPS against SNR.\label{5}}
	\end{figure*}
	\begin{figure*}
		\centering
	\begin{minipage}[tbp]{0.3\linewidth}
		\centering
		\subfloat[]{
			\label{PSNR_CBR_AWGN}
			\centering
			\includegraphics[width=1\textwidth]{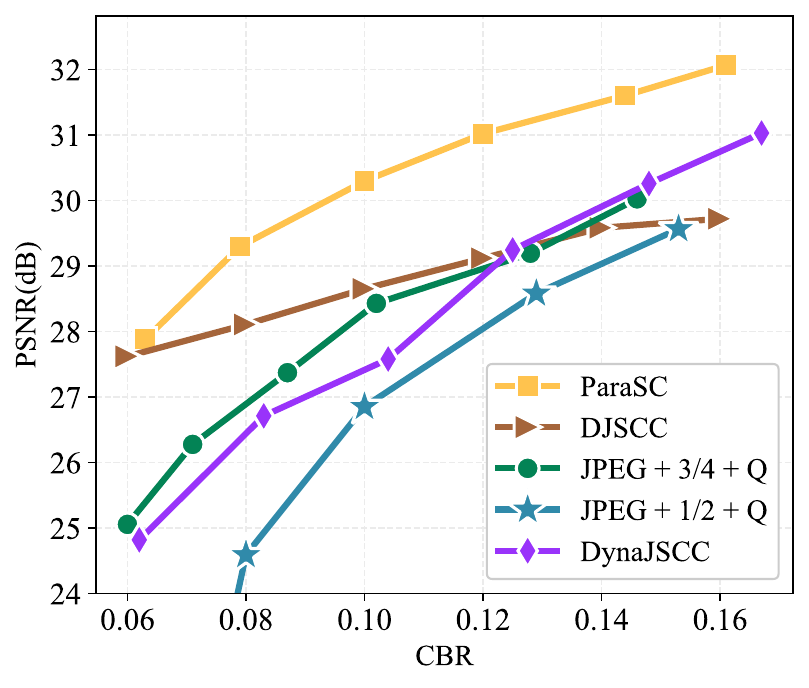}
		}
	\end{minipage}
	\begin{minipage}[tbp]{0.3\linewidth}
		\centering
		\subfloat[]{
			\label{SSIM_CBR_AWGN}
			\centering
			\includegraphics[width=1\textwidth]{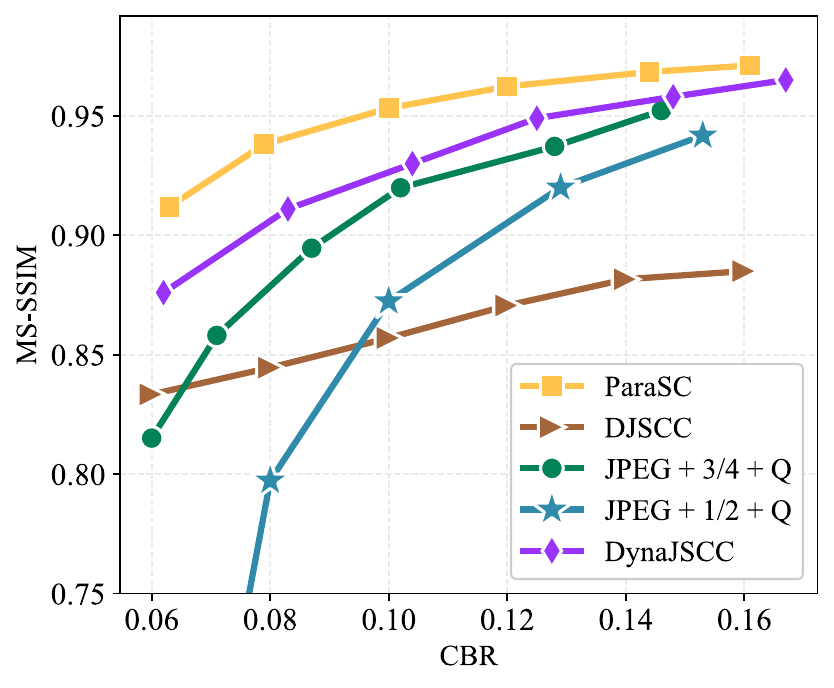}}
	\end{minipage}
	\begin{minipage}[tbp]{0.3\linewidth}
		\centering
		\subfloat[]{
			\label{LPIPS_CBR_AWGN}
			\centering
			\includegraphics[width=1\textwidth]{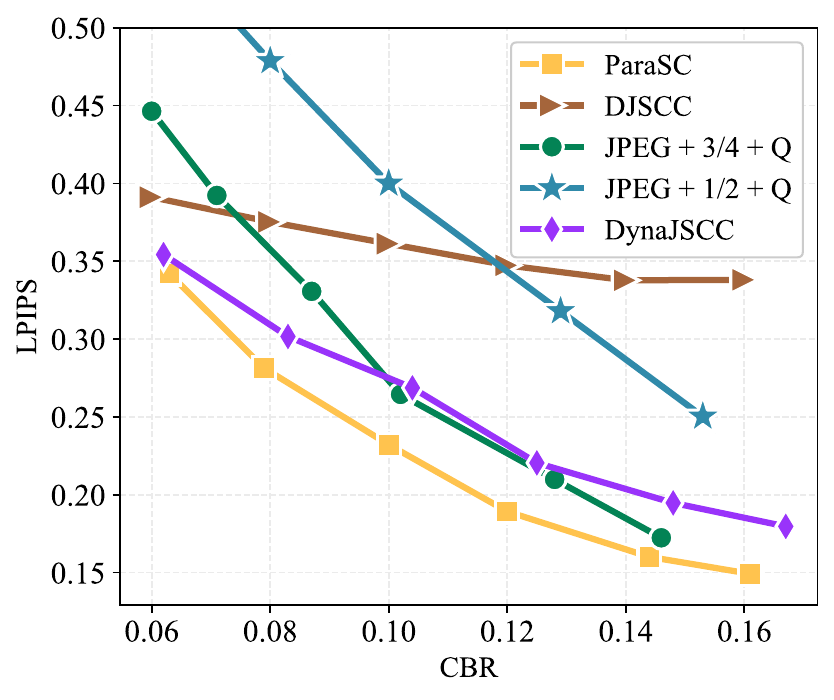}}
	\end{minipage}
	\caption{Comparisons of reconstruction results on CVRG-Pano dataset under AWGN channel in terms of ParaSC and other baselines with SNR=$6$dB; (a) PSNR against CBR; (b) MS-SSIM against CBR; (c) LPIPS against CBR.\label{6}}
\end{figure*}
\subsubsection{Baseline Schemes}
We compare our proposed ParaSC with two types of baseline schemes: conventional SSCC and DL-based JSCC. 
\begin{itemize}
	\item For SSCC, we use JPEG compression with LDPC channel coding and QPSK modulation. Here, we consider LDPCs with code rates $0.75$, $0.66$ and $0.5$, which are denoted by "JPEG + 3/4 + Q", "JPEG + 2/3 + Q" and "JPEG + 1/2 + Q", respectively\footnote{For brevity we use '+' to represent the combination of source, channel coding, and modulation schemes}. 
	\item For the DL-based end-to-end coding framework, we consider DJSCC~\cite{djscc}, a pioneering framework of JSCC based on autoencoder. We set the SNR to $10$ dB for training and vary the number of filters in the last convolution layer of DJSCC's encoder to adjust its CBR. 
	\item Since there exists rate-adaptive mechanism in ParaSC, we also compare with DynaJSCC~\cite{dynamicjscc}, an SNR-based rate adaptive JSCC. Here, we set $G_s=9, G_n=2$, so the achievable CBR region yields $\left[0.042,0.229\right]$ in order to cover a sufficiently wide range of transmission rates. DynaJSCC is trained by various channel SNRs ranging from 2dB to 12dB.
\end{itemize}

\subsubsection{Evaluation Details}
As for image quality metrics, we first choose peak signal-to-noise (PSNR), which is commonly adopted in various researches to evaluate the per-pixel difference between images. The PSNR between two images $\bm{x}$ and $\hat{\bm{x}}$ is defined as
\begin{equation}
	\mbox{PSNR}=10\log_{10}{\frac{\mbox{MAX}^2}{\mbox{MSE}}}\notag,
\end{equation}
where \mbox{MAX} denotes the possible peak value of the image pixel. Since the images we use have $8$ bits per pixel, we have $\mbox{MAX}=2^8-1=255$. Despite the simplicity and effectiveness, PSNR does not align with human's visual perception in many cases. Therefore, to conduct a more comprehensive assessment, we further include MS-SSIM~\cite{msssim} and LPIPS~\cite{lpips} as two perceptual quality metrics.

\subsection{Performance Evaluation on AWGN Channel}
\begin{figure*}
	\centering
	\begin{minipage}[tbp]{0.3\linewidth}
		\centering
		\subfloat[]{
			\label{PSNR_Rayleigh_SNR}
			\centering
			\includegraphics[width=1\textwidth]{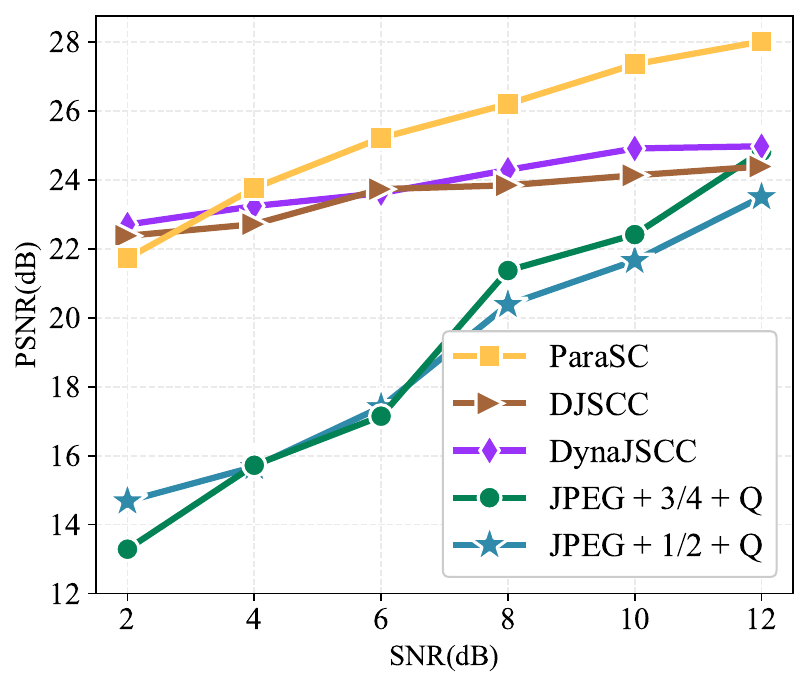}
		}
	\end{minipage}
	\begin{minipage}[tbp]{0.3\linewidth}
		\centering
		\subfloat[]{
			\label{SSIM_Rayleigh_SNR}
			\centering
			\includegraphics[width=1\textwidth]{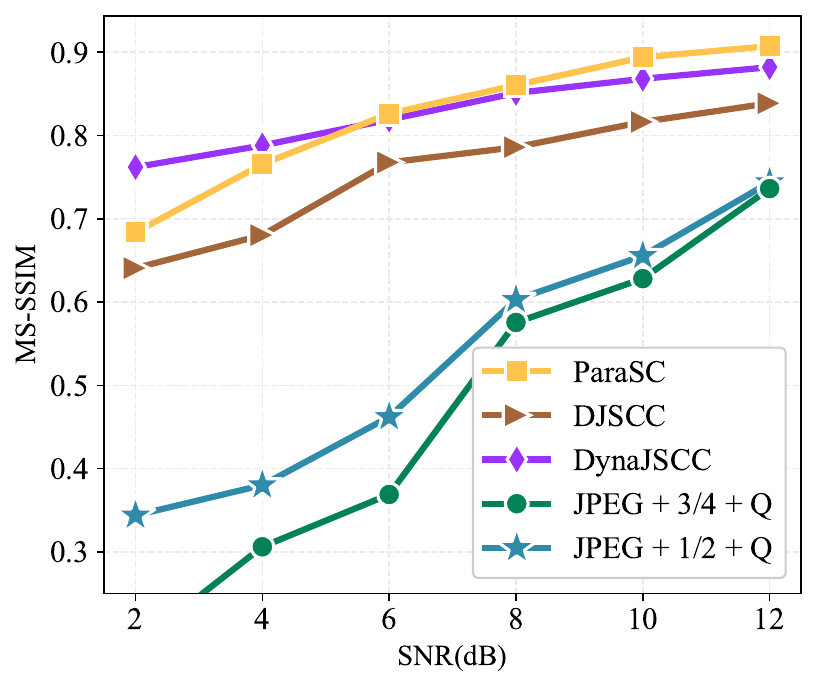}
		}
	\end{minipage}
	\begin{minipage}[tbp]{0.3\linewidth}
		\centering
		\subfloat[]{
			\label{LPIPS_Rayleigh_SNR}
			\centering
			\includegraphics[width=1\textwidth]{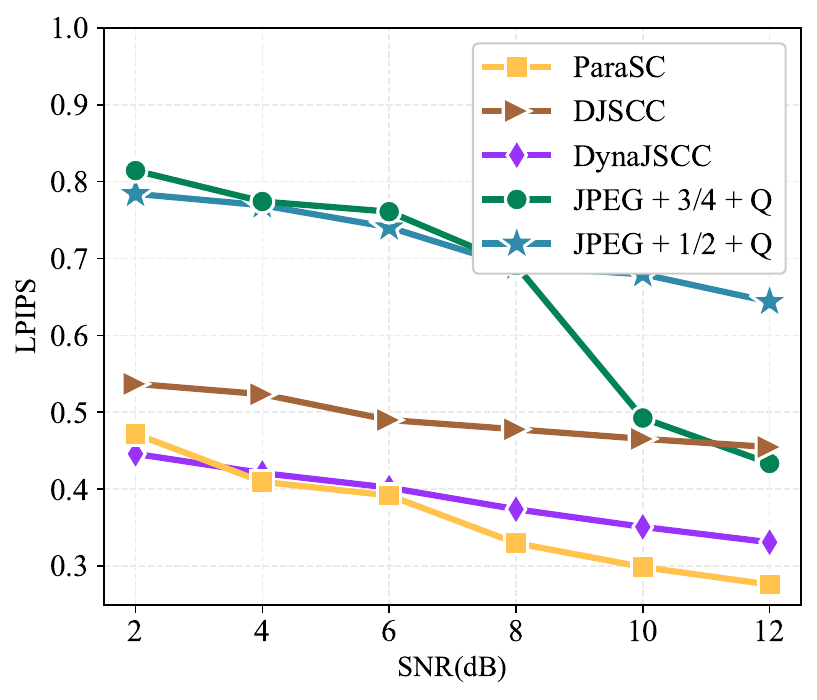}
		}
	\end{minipage}
	\caption{Comparison of reconstruction results of ParaSC and other baselines on CVRG-Pano dataset under Rayleigh channel with CBR=$1/12$. (a) PSNR against SNR; (b) MS-SSIM against SNR. (c) LPIPS against SNR. \label{figure:rayleigh}}
\end{figure*}
\begin{figure*}
	\centering
	\begin{minipage}[tbp]{0.3\linewidth}
		\centering
		\subfloat[]{
			\label{PSNR_Rayleigh_SNR_kodak}
			\centering
			\includegraphics[width=1\textwidth]{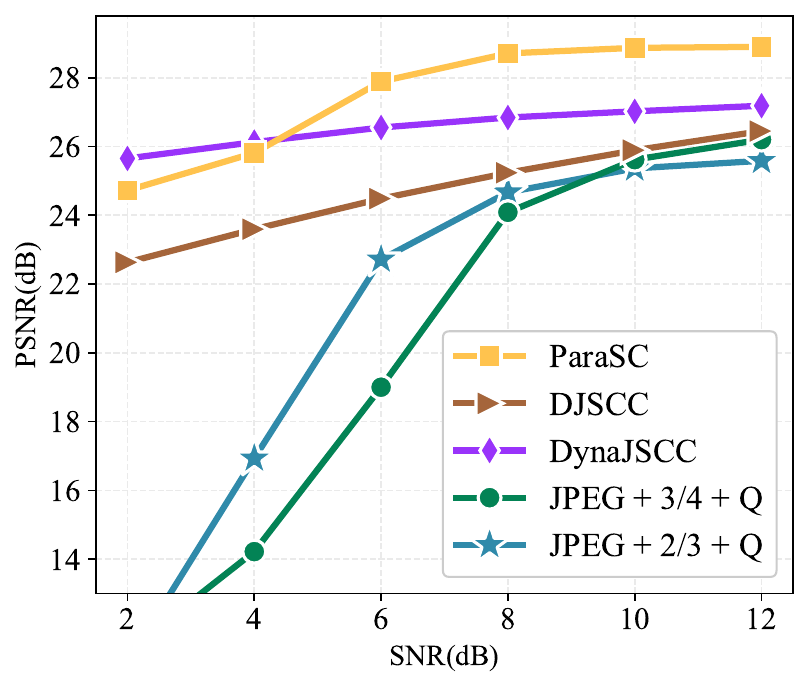}
		}
	\end{minipage}
	\begin{minipage}[tbp]{0.3\linewidth}
		\centering
		\subfloat[]{
			\label{SSIM_Rayleigh_SNR_kodak}
			\centering
			\includegraphics[width=1\textwidth]{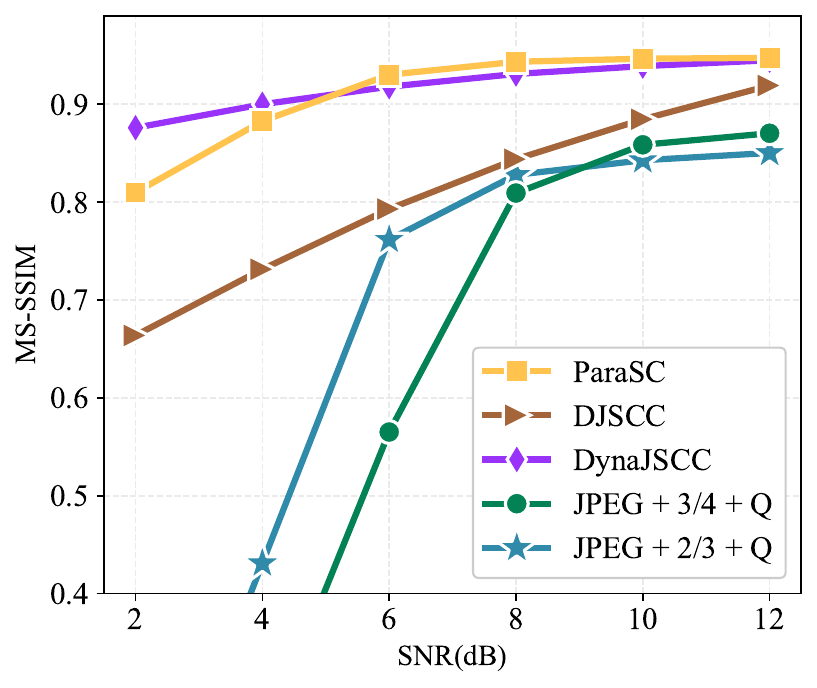}
		}
	\end{minipage}
	\begin{minipage}[tbp]{0.3\linewidth}
		\centering
		\subfloat[]{
			\label{LPIPS_Rayleigh_SNR_kodak}
			\centering
			\includegraphics[width=1\textwidth]{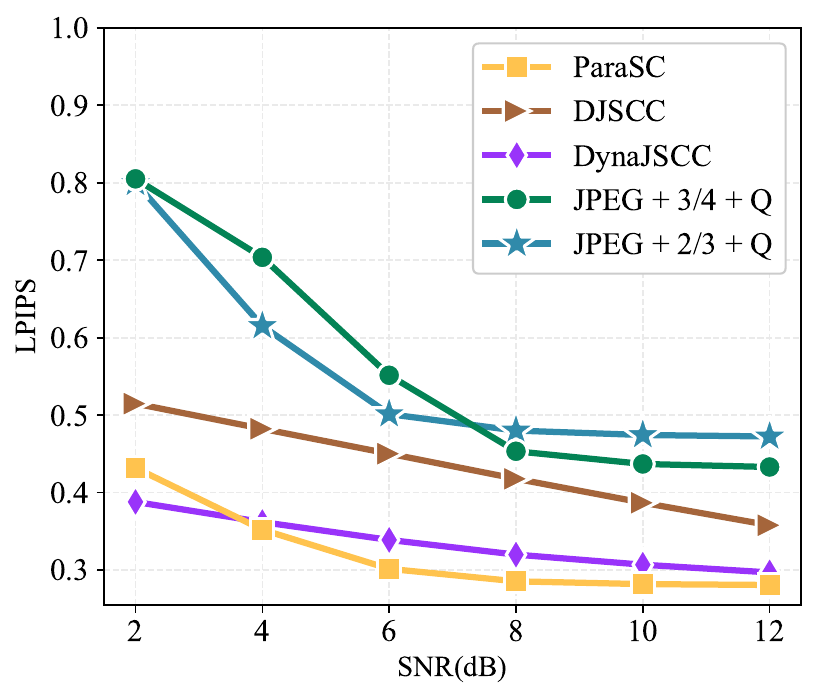}
		}
	\end{minipage}
	\caption{Comparison of reconstruction results of ParaSC and other baselines on Kodak24 dataset under Rayleigh channel with CBR=$1/12$. (a) PSNR against SNR; (b) MS-SSIM against SNR. (c) LPIPS against SNR. \label{figure:rayleigh_kodak}}
\end{figure*}
First, we investigate the image reconstruction quality of our proposed ParaSC along with that of the other baseline schemes for AWGN channels. In the following experiments, rectangular markers represent our framework, triangular markers denote DJSCC networks and diamond ones denote DynaJSCC network, while circular and star ones stand for JPEG-based SSCC with different LDPC code rates, respectively. In Fig.~\ref{5}\subref{PSNR_SNR_AWGN}-\subref{LPIPS_SNR_AWGN}, the PSNR, MS-SSIM and LPIPS are plotted as functions of the SNR for a fixed overall CBR=$0.1$. Clearly, the JPEG methods suffer from the 'cliff effect'. E.g., for SNR$\ge$$6$dB, the performance of "JPEG + 3/4 + Q" is steady and decent but for SNR $<$$6$dB, it drops drastically. Under this circumstance, ParaSC outperforms both the conventional methods and DJSCC across various SNRs in terms of different evaluation metrics. Provided with enough channel quality (SNR$\ge$$6$dB), the performance of ParaSC reaches a plateau since the transmission is dominated by the conventional image stream. It is also noticeable that ParaSC performs better than JPEG method due to the residual compensation. For SNR$<$$6$dB, the corrupted JPEG image has some impact on the performance of ParaSC, resulting in its PSNR and MS-SSIM being slightly inferior to DynaJSCC at SNR$=$$2$dB. However, the 'cliff effect' phenomenon is alleviated in ParaSC. This is attributed to our PAGNet which inhibits the image stream under a low SNR environment by assigning higher aggregation weights to the semantic stream.

Then we investigate the rate-distortion performance of several baselines under AWGN channel. The CBR of ParaSC is varied by changing the JPEG quality factor $q$. The optimal CBR learned by the policy network in DynaJSCC is 0.167, with other CBR points achieved by manually changing its binary mask vector to obtain a multiple-rate single model in the test stage. Figs.~\ref{6}\subref{PSNR_CBR_AWGN}-\subref{LPIPS_CBR_AWGN} show the PSNR, MS-SSIM and LPIPS against CBR among all competitors, respectively, by fixing the channel quality SNR=$6$dB. For all the evaluation metrics, ParaSC shows an overwhelming advantage across the whole CBR region. It is seen that the performance of JPEG-built methods drops slightly faster when the CBR is around 0.06 because this is where the JPEG almost reaches its compression limit on CVRG-Pano dataset. However, with the increase of CBR, a dramatic coding gain is immediately observed in ParaSC and JPEG methods. In contrast, although transmitting more contents, DJSCC becomes stagnant in the performance and is finally overtaken by JPEG methods. Meanwhile, ParaSC can greatly benefit from increasing CBR due to its built-in conventional image codec. Note that DynaJSCC underperforms at the low CBRs ($\le 0.1$) because this point is far from its optimum (0.167) and its corresponding model weights are rarely activated during the training stage. However, as the CBR approaches the optimum, DynaJSCC's PSNR stands out from DJSCC and conventional codec, with its MS-SSIM and LPIPS consistently beating them across the CBR region. Compared with DynaJSCC, ParaSC saves up to 28.14\% CBR to reach around 31dB in PSNR.

To conclude, in AWGN channel, ParaSC harvests a comprehensive performance on diverse assessment metrics including machine-level distortion and human-level perception, while overcoming the cliff effect of conventional codec.

\subsection{Performance Evaluation on Rayleigh Channel}
\begin{figure*}[tbp]
	\centering
	\begin{minipage}[t]{0.07\linewidth}
		\centering
		\vspace*{9ex} 
		\textbf{$q=5$}
		
		\vspace*{14ex} 
		\textbf{$q=10$}
		
		\vspace*{15ex} 
		\textbf{$q=30$}
	\end{minipage}
	\begin{minipage}[t]{0.263\linewidth}
		\centering
		\subfloat{\includegraphics[width=1\textwidth]{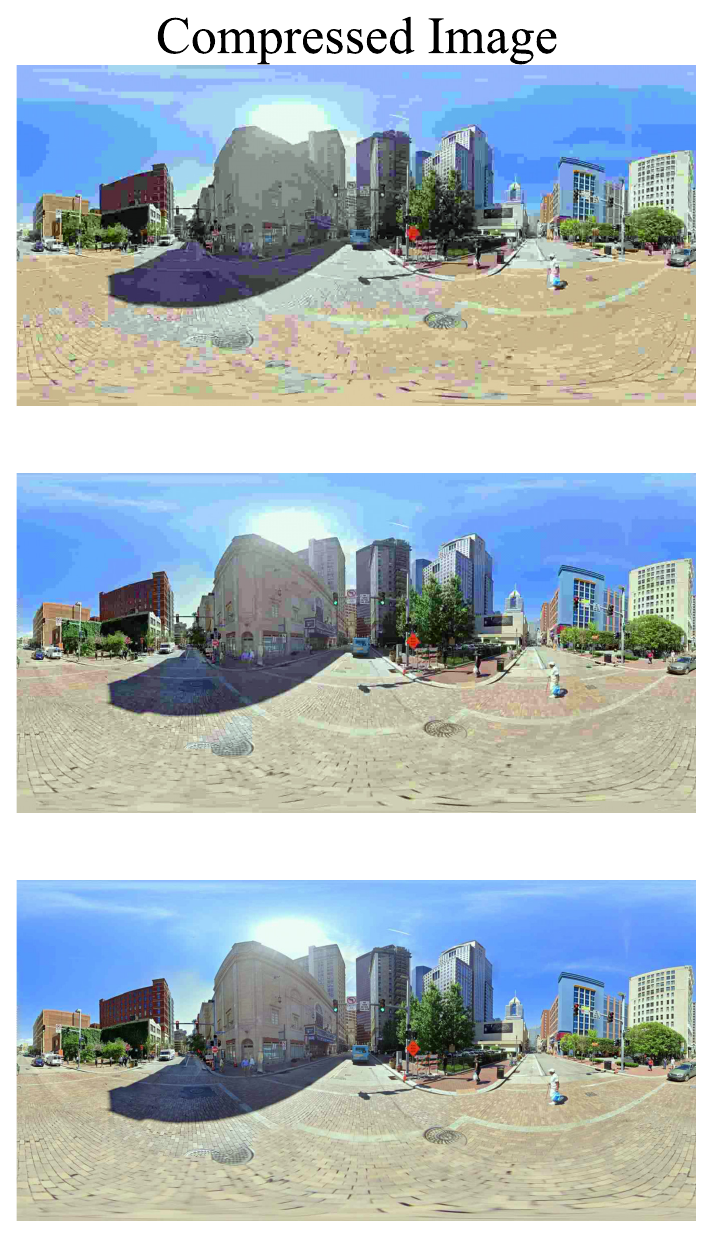}}
	\end{minipage}
	\begin{minipage}[t]{0.32\linewidth}
		\centering
		\subfloat{\includegraphics[width=1\textwidth]{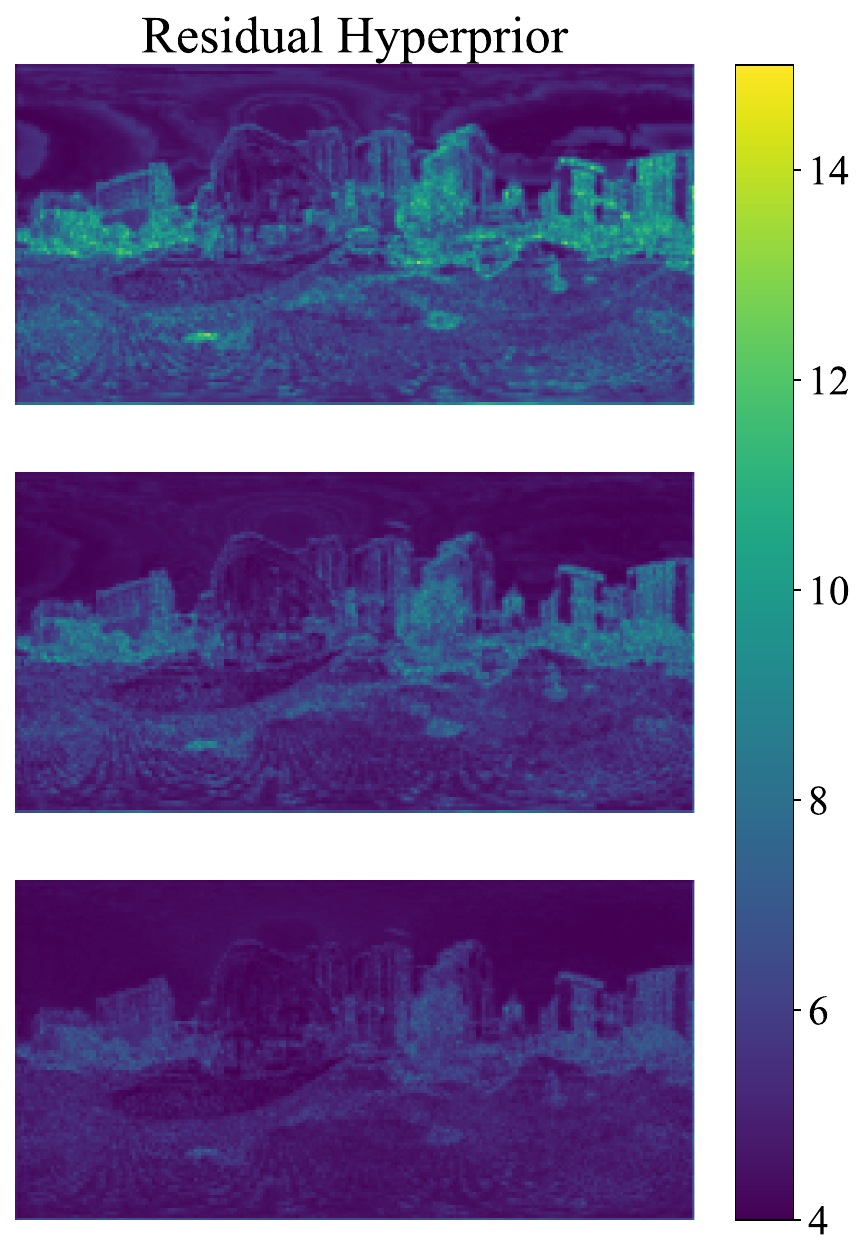}}
	\end{minipage}
	\begin{minipage}[t]{0.32\linewidth}
		\centering
		\subfloat{\includegraphics[width=1\textwidth]{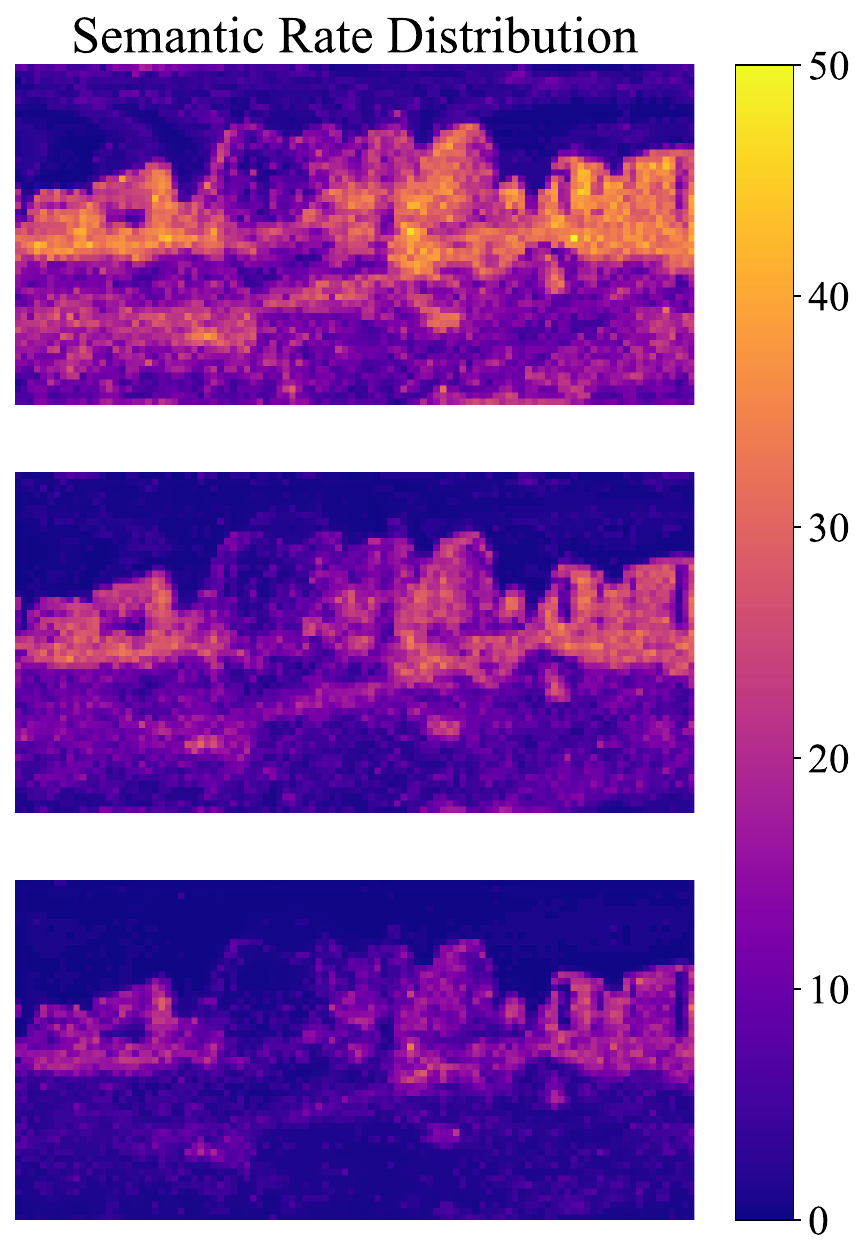}}
	\end{minipage}
	
	\caption{The rate adaptation results under different residual conditions, which are controlled by conventional compression quality factor $q\in\{5,10,30\}$. A larger $q$ results in the image with a higher quality.}
	\label{fig:rate adaptation visual}
\end{figure*}
In this subsection, we explore the reconstruction performance of ParaSC in comparison with baseline schemes under a Rayleigh block fading channel. In Fig.~\ref{figure:rayleigh}\subref{PSNR_Rayleigh_SNR}-\subref{LPIPS_Rayleigh_SNR}, the achieved PSNR, MS-SSIM and LPIPS of ParaSC and other baselines schemes on CVRG-Pano are presented over SNR by fixing CBR=$1/12$, respectively. It can be observed that in PSNR and LPIPS metrics, our architecture surpasses both DL-based and conventional transmission schemes in a wide range of SNR (4-12dB). Due to the broken format during Huffman decoding, the JPEG methods are poor in coping with channel fading. As two streams are coupled with each other and PAGNet cannot fully decouple them, the performance of ParaSC is slightly lower than DynaJSCC under low-SNR fading channel environments. With the improvement of channel quality, ParaSC becomes superior to the baseline schemes and further expands its gain. Furthermore, we switch to the Kodak24 dataset and present the reconstruction results under the Rayleigh fading channel in Fig.~\ref{figure:rayleigh_kodak}. As shown, ParaSC exhibits competitive performance compared to DynaJSCC on MS-SSIM and superior performance on PSNR and LPIPS across most SNR conditions, demonstrating its generalization capability.

\subsection{Ablation Study}
In this section, we dive deep into the intrinsic properties of ParaSC by studying the conditional rate adaptation mechanism. Experiments reveal the positive interaction between the image stream and the semantic stream, which enables our framework to improve the single-stream SSCC's performance in exchange for an additional but trivial semantic transmission. In a broad range of scenarios, ParaSC proves to maintain lightweight in the overhead of the semantic stream.

First, Fig.~\ref{fig:rate adaptation visual} visualizes the residual hyperpriors and the adapted semantic rate distributions given the compressed JPEG images with different quality factors. Clearly, the learned conditional entropy model assigns different code lengths to different regions of the semantic features. Besides, the increase of the JPEG coding quality causes a decrease of the residual, which ultimately leads to the reduction of semantic code length. For instance, the compressed image with $q=5$ is blurry in the background (sky) region, resulting in a certain amount of rate to transmit the corresponding region in the semantics; however, when $q$ grows to 30, JPEG can compress the background with little distortion, which means conventional coding is almost lossless in this region so the allocated semantic rate for the background approaches zero. 

\begin{figure}[htbp]
	\centering
	\includegraphics[width=0.4\textwidth]{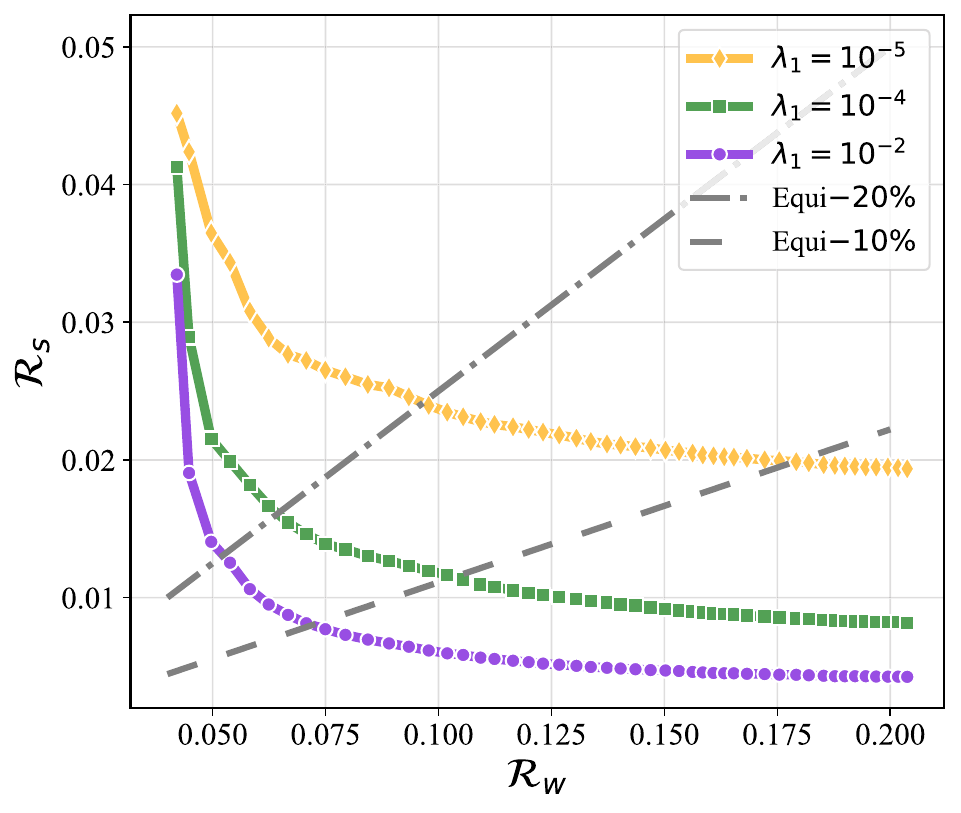}
	\caption{Trade-off between the image stream rate $\mathcal{R}_w$ and the semantic stream rate $\mathcal{R}_s$ trained under different $\lambda_1$. All the points on the line "Equi$-\delta\%$" satisfies $\frac{\mathcal{R}_s}{\mathcal{R}_w+\mathcal{R}_s}=\delta\%$. }
	\label{fig:zw-zs}
\end{figure}

We offer a quantitative analysis of the conditional rate adaptation in Fig.~\ref{fig:zw-zs}, where the image stream rate $\mathcal{R}_w$ is controlled by varing the quality factor $q$. A strong negative correlation between $\mathcal{R}_w$ and $\mathcal{R}_s$ can be observed, indicating that the semantic stream is complementary to the image stream. Different trade-offs can be achieved by tuning the rate coefficient $\lambda_1$ in the loss function, which extends our framework to the scenarios with limited bandwidth budgets. We also plot the equi-proportion lines satisfying $\frac{\mathcal{R}_s}{\mathcal{R}_w+\mathcal{R}_s}=0.1, 0.2$ to show the transmission bandwidth occupied by semantic stream. For $\lambda_1=0.01$, when $\mathcal{R}_w>0.07$, the proportion of $\mathcal{R}_s$ in the overall rate is narrowed down to $\le10\%$, which proves that semantic stream is lightweight and induces negligible incremental communication cost at the transmitter, thus ensuring our framework is compatible to practical SSCC.
\begin{table}[bp]
	\renewcommand\arraystretch{1.1}
	\begin{center}
		\caption{Comparisons in terms of the complexity of JSCC transmitters. The throughput is measured on a single NVIDIA RTX 3090 GPU, with the images cropped to $512\times1024$.\label{Tab3}}
		\begin{tabular}{c|ccc}
			\toprule
			\multirow{2}{*}{Method} & Parameters & FLOPs & Throughput \\
			& (million) & (G) & (images / sec) \\
			\midrule
			DJSCC & 0.1 & 1.55 & 856.25 \\	
			DynaJSCC & 3.3 & 64.37 & 78.26\\
			WITT & 14.1 & 359.38 & 24.31 \\
			\midrule
			ParaSC (\textbf{Ours}) & 2.3 & 45.64 & 142.13 \\
			\bottomrule
		\end{tabular}
	\end{center}
\end{table}

\begin{figure}[tbp]
	\centering
	\includegraphics[width=0.45\textwidth]{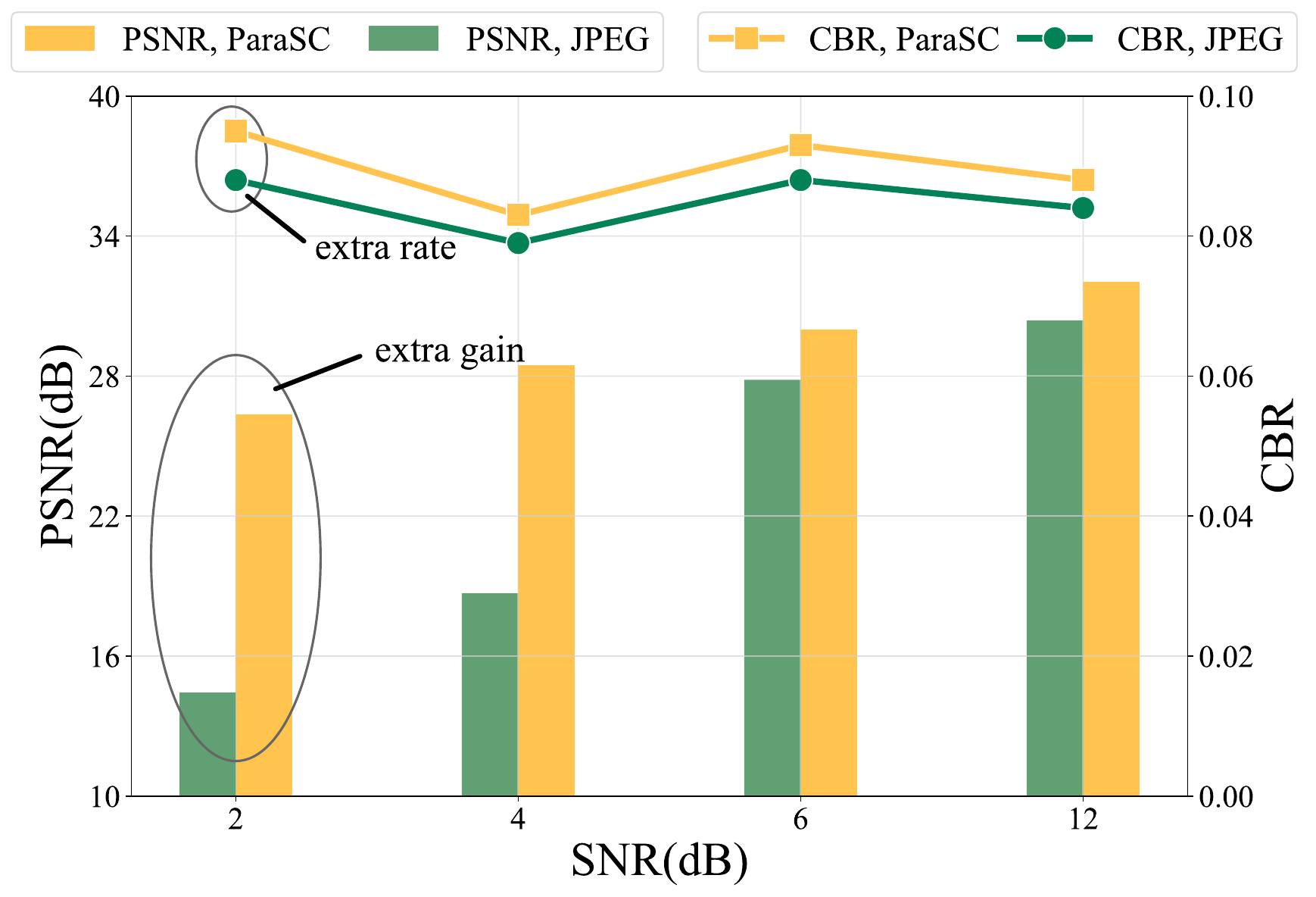}
	\caption{Comparisons between ParaSC and the single conventional branch of ParaSC ("JPEG") in terms of the reconstruction quality and bandwidth cost under varied SNRs.}
	\label{fig:barline}
\end{figure}

To further investigate the performance variation caused by rate adaptation, we remove the semantic stream from ParaSC and get a single-stream JPEG-based coding method, then studying the performance gap between this single-stream and our parallel-stream framework. Fig.~\ref{fig:barline} shows the PSNR and CBR of these two methods under different SNRs. Note that the coding process of "JPEG" is identical to the conventional coding in "ParaSC", which means they adopt the same non-learnable parameters in SSCC. Hence the only difference is the participation of semantic stream. It is shown that our semantic stream is small in quantity but large in effect. Continuous PSNR gain is observed across the SNR interval, with the largest gain obtained at the worst channel environment. When SNR$=2$dB, transmitting extra $8.3\%$ rate of semantic stream brings $82.4\%$ PSNR gain. This is also attributed to the PAGNet which filters the distortion patches in the compressed image $\bm{\hat{x}}_c$ and lets the semantics dominate the image generation process. Without PAGNet, even if the conditional entropy model compresses the semantics to the optimal rate, the fluctuating channel noise will distort $\bm{\hat{x}}_c$ and thus ruin the condition information for semantic decoder to rely on.
\begin{figure*}[htbp]
	\begin{minipage}[t]{0.12\linewidth}
		\centering
		\vspace*{-9ex} 
		\textbf{Original}
		
		\vspace*{16ex} 
		\textbf{ParaSC}
		
		\vspace*{16ex} 
		\textbf{DynaJSCC}
		
		\vspace*{15ex} 
		\textbf{DJSCC}
		
		\vspace*{15ex} 
		\textbf{JPEG}
	\end{minipage}
	\begin{minipage}[t]{0.44\linewidth}
		\centering
		\begin{minipage}[t]{1\linewidth}
			\centering
			\includegraphics[width=0.6\textwidth]{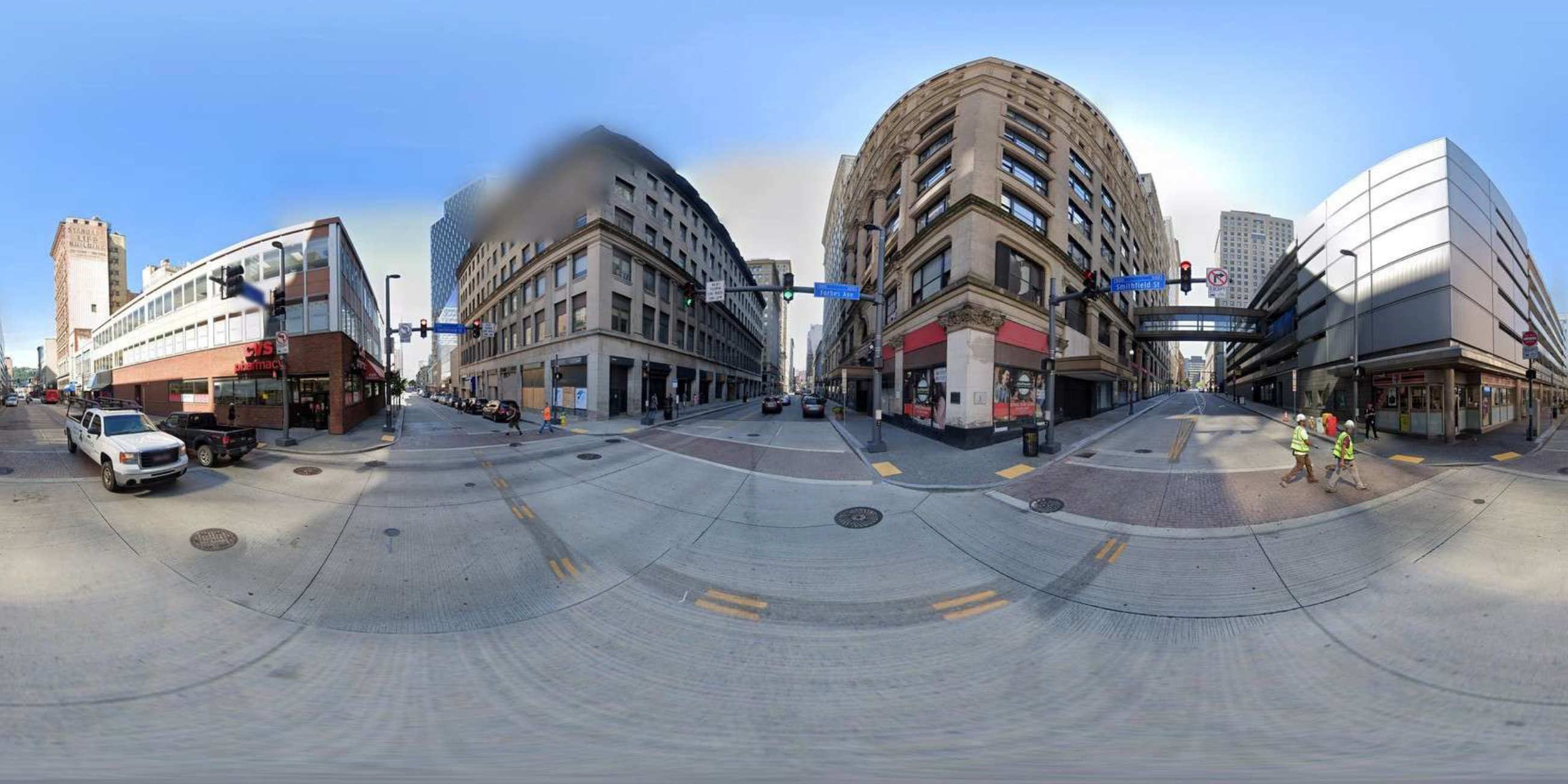}
			\includegraphics[width=0.198\textwidth]{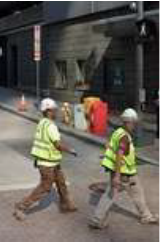}
			\centering SNR$=2$dB
			\vspace{0.05cm}
		\end{minipage}
		
		\begin{minipage}[t]{1\linewidth}
			\centering
			\includegraphics[width=0.6\textwidth]{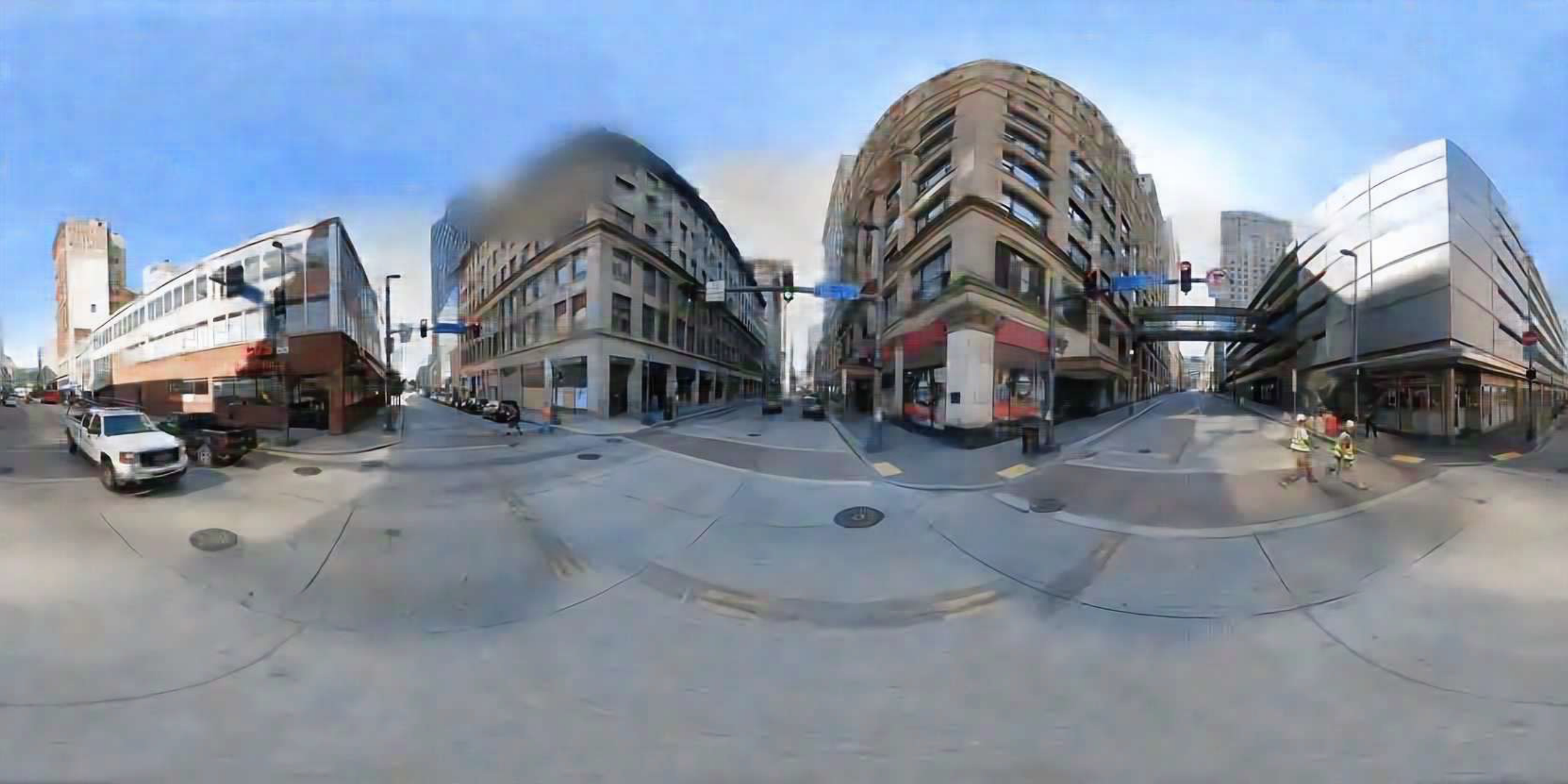}
			\includegraphics[width=0.198\textwidth]{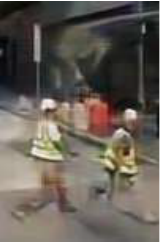}
			\centering CBR$=0.098$\quad PSNR$=26.31$
			\vspace{0.05cm}
		\end{minipage}
		
		\begin{minipage}[t]{1\linewidth}
			\centering
			\includegraphics[width=0.6\textwidth]{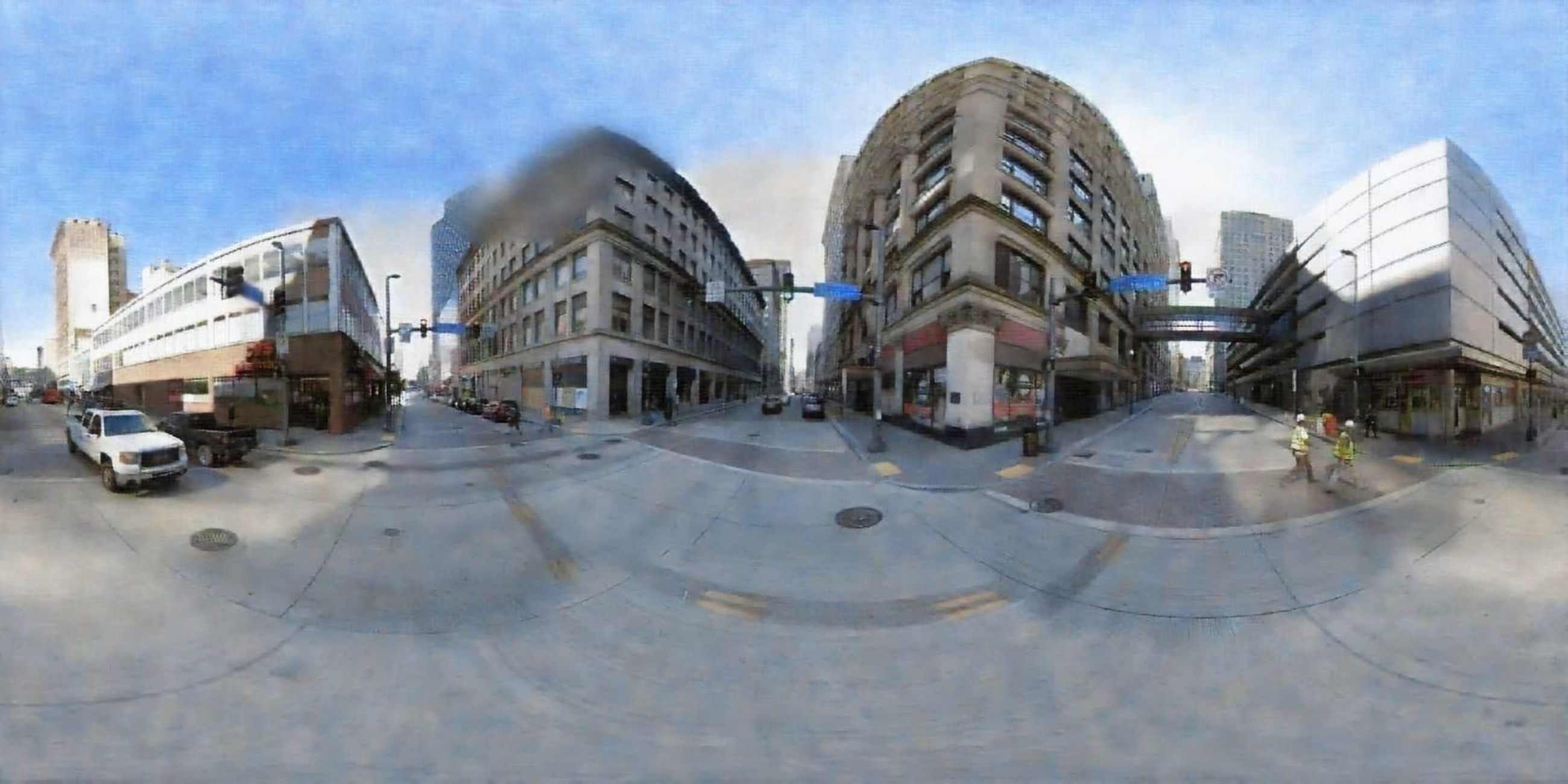}
			\includegraphics[width=0.198\textwidth]{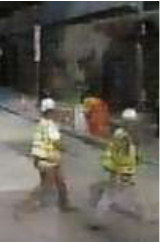}
			\centering CBR$=0.104$\quad PSNR$=25.81$
			\vspace{0.05cm}
		\end{minipage}
		
		\begin{minipage}[t]{1\linewidth}
			\centering
			\includegraphics[width=0.6\textwidth]{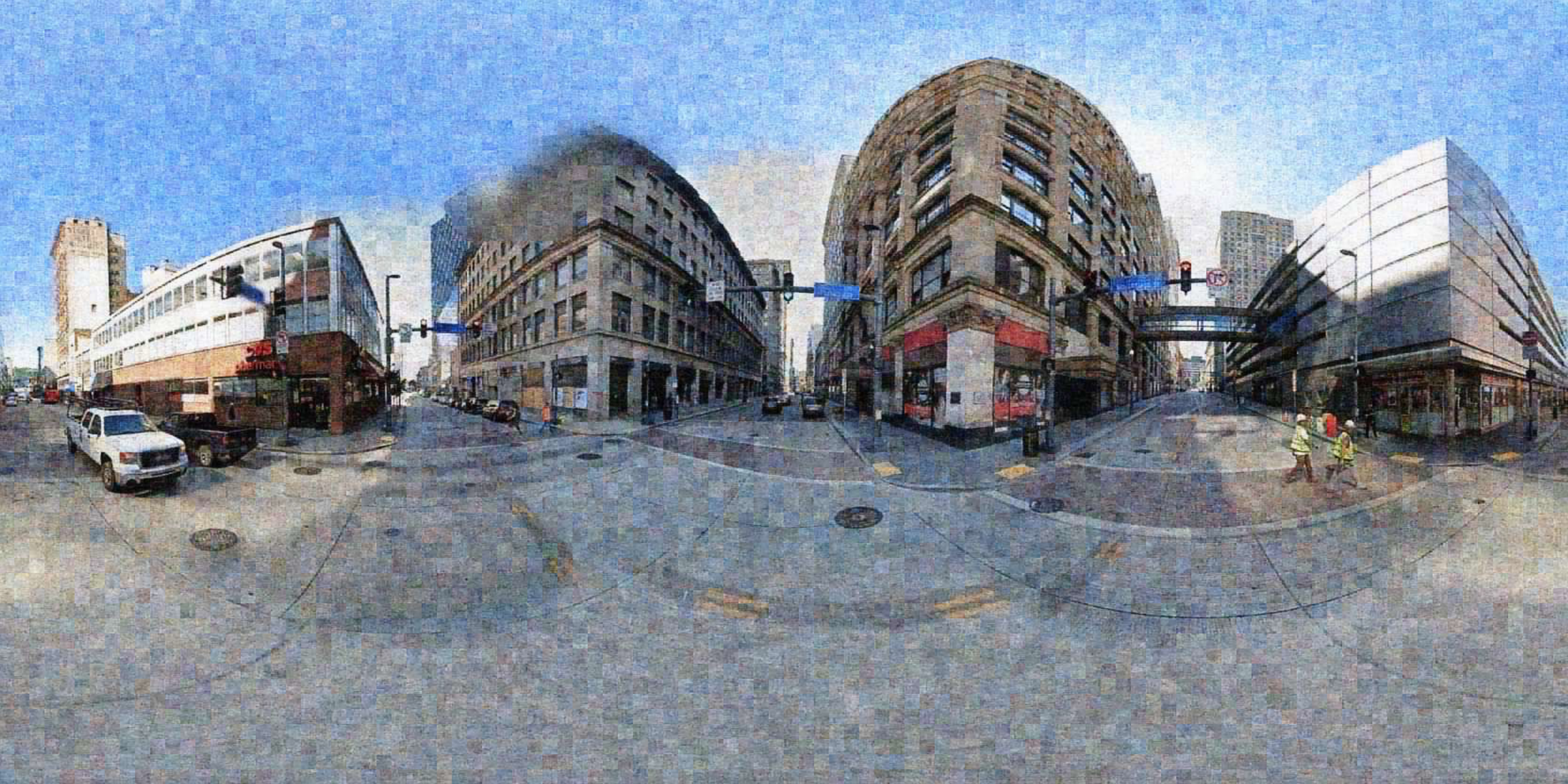}
			\includegraphics[width=0.198\textwidth]{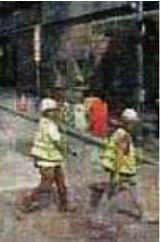}
			\centering CBR$=0.100$\quad PSNR$=24.26$
			\vspace{0.05cm}
		\end{minipage}
		
		\begin{minipage}[t]{1\linewidth}
			\centering	
			\includegraphics[width=0.6\textwidth]{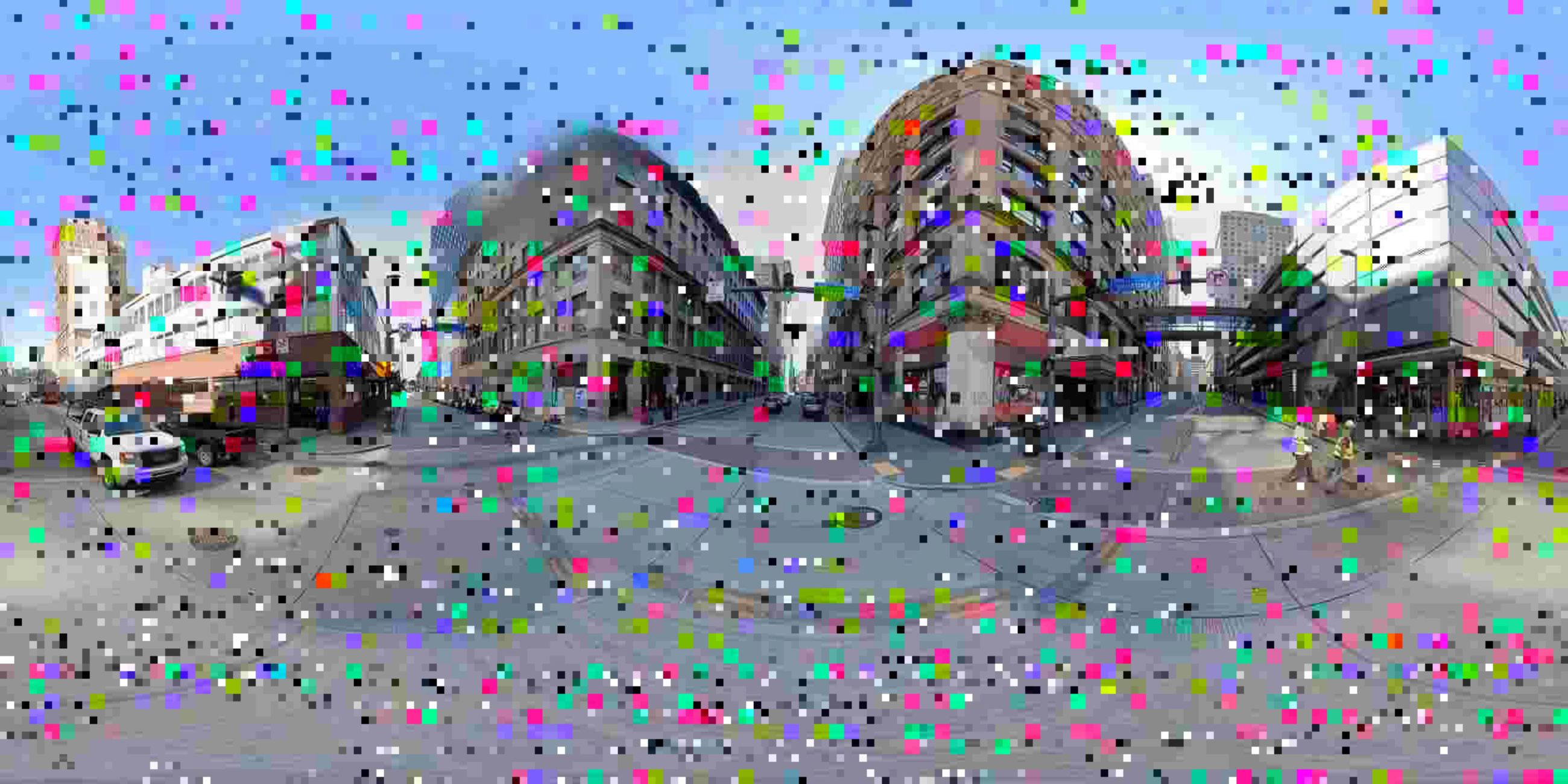}
			\includegraphics[width=0.198\textwidth]{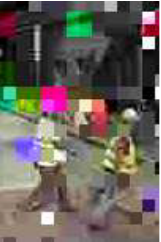}
			\centering CBR$=0.107$\quad PSNR$=15.96$
			\vspace{0.05cm}
		\end{minipage}	
	\end{minipage}
	\begin{minipage}[t]{0.44\linewidth}
		\centering
		\includegraphics[width=0.6\textwidth]{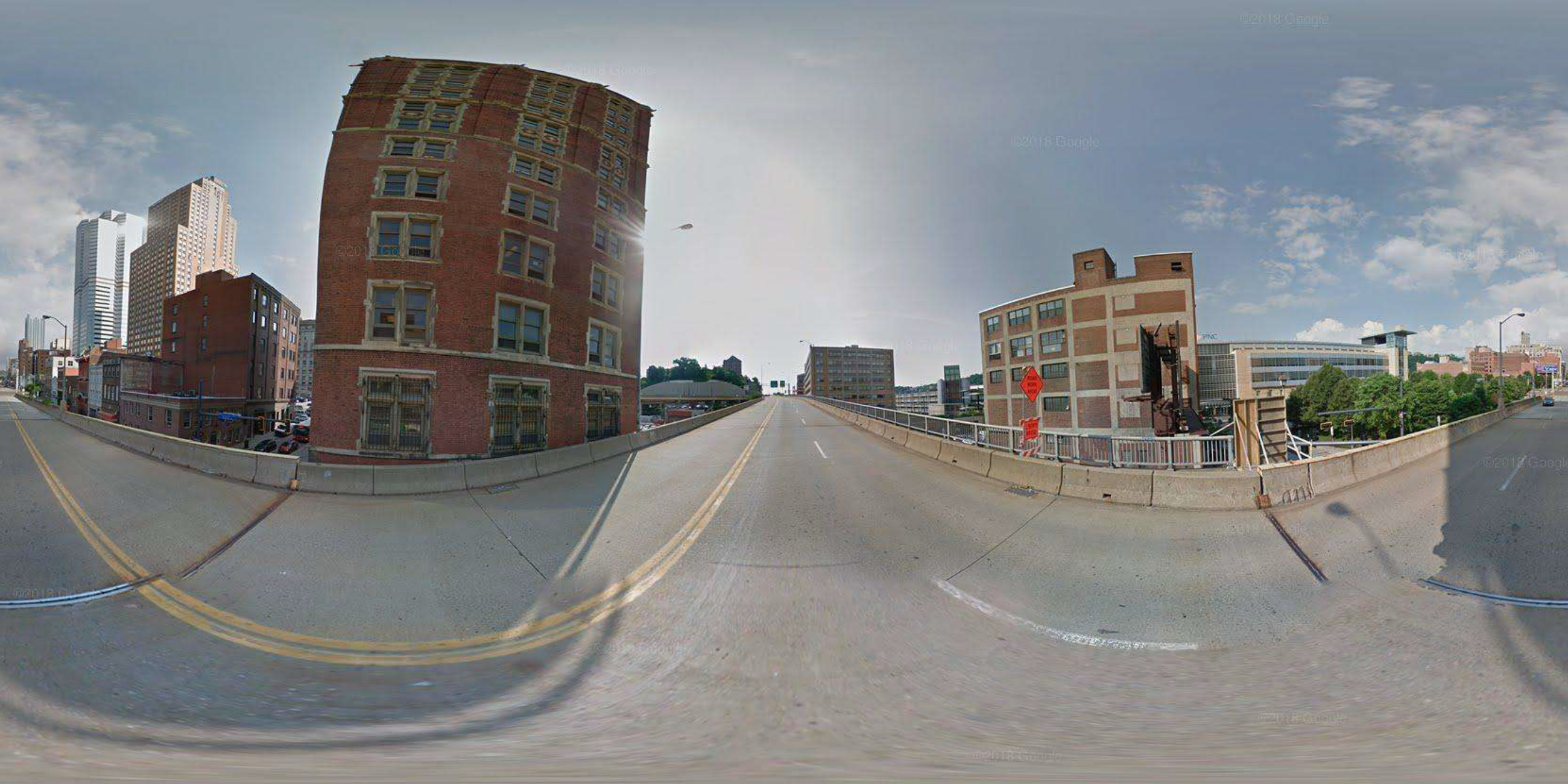}
		\includegraphics[width=0.198\textwidth]{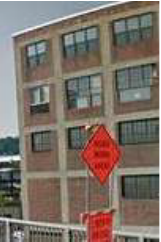}
		\centering SNR$=6$dB
		\vspace{0.05cm}
		
		\includegraphics[width=0.6\textwidth]{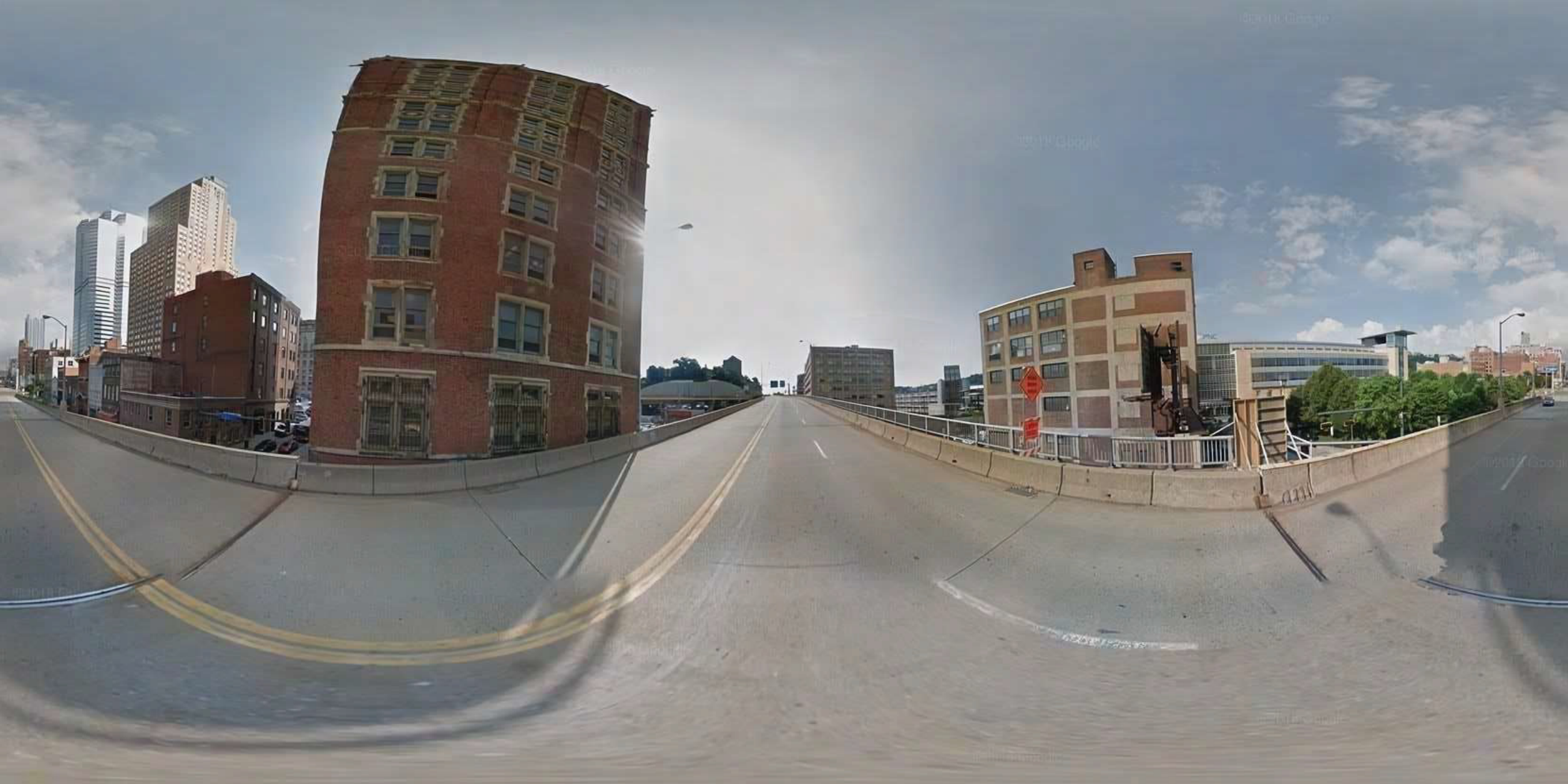}
		\includegraphics[width=0.198\textwidth]{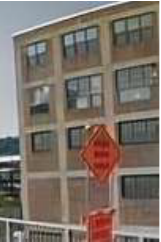}
		\centering CBR$=0.133$\quad PSNR$=34.04$
		\vspace{0.05cm}
		
		\includegraphics[width=0.6\textwidth]{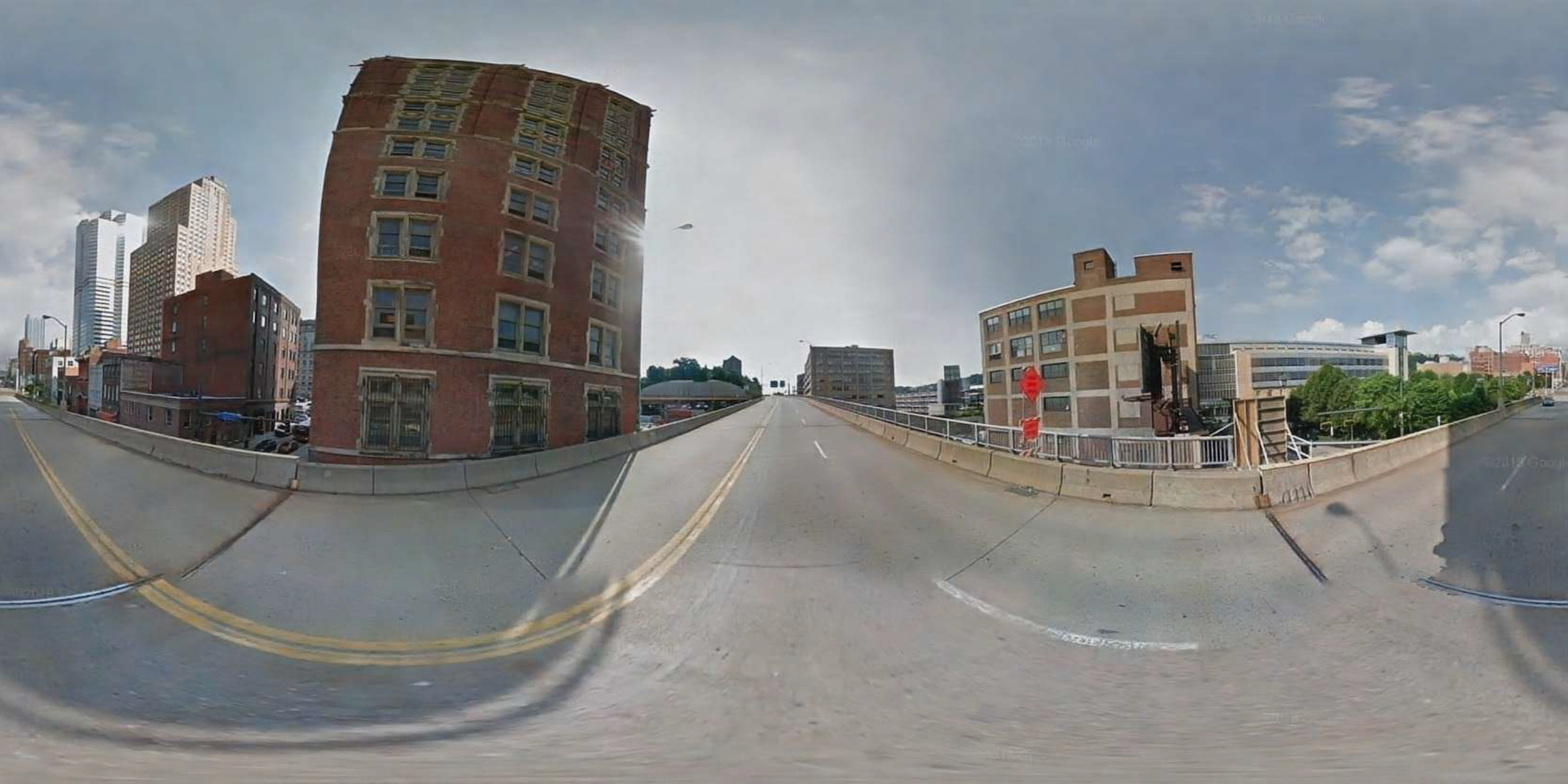}
		\includegraphics[width=0.198\textwidth]{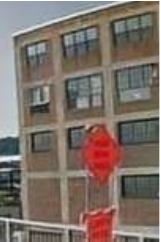}
		\centering CBR$=0.167$\quad PSNR$=33.34$
		\vspace{0.05cm}
		
		\includegraphics[width=0.6\textwidth]{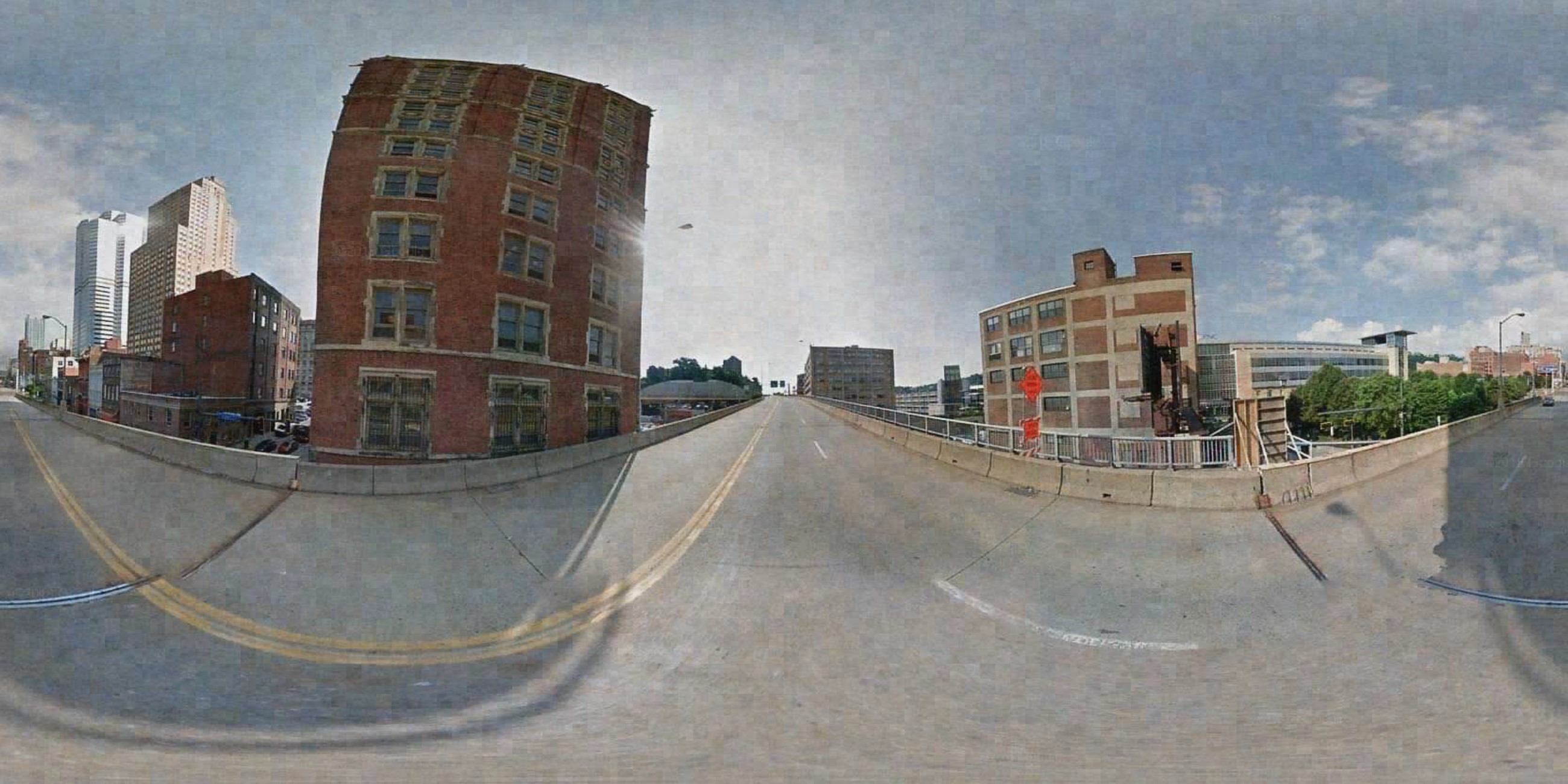}
		\includegraphics[width=0.198\textwidth]{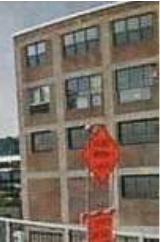}
		\centering CBR$=0.160$\quad PSNR$=32.03$
		\vspace{0.05cm}
		
		\includegraphics[width=0.6\textwidth]{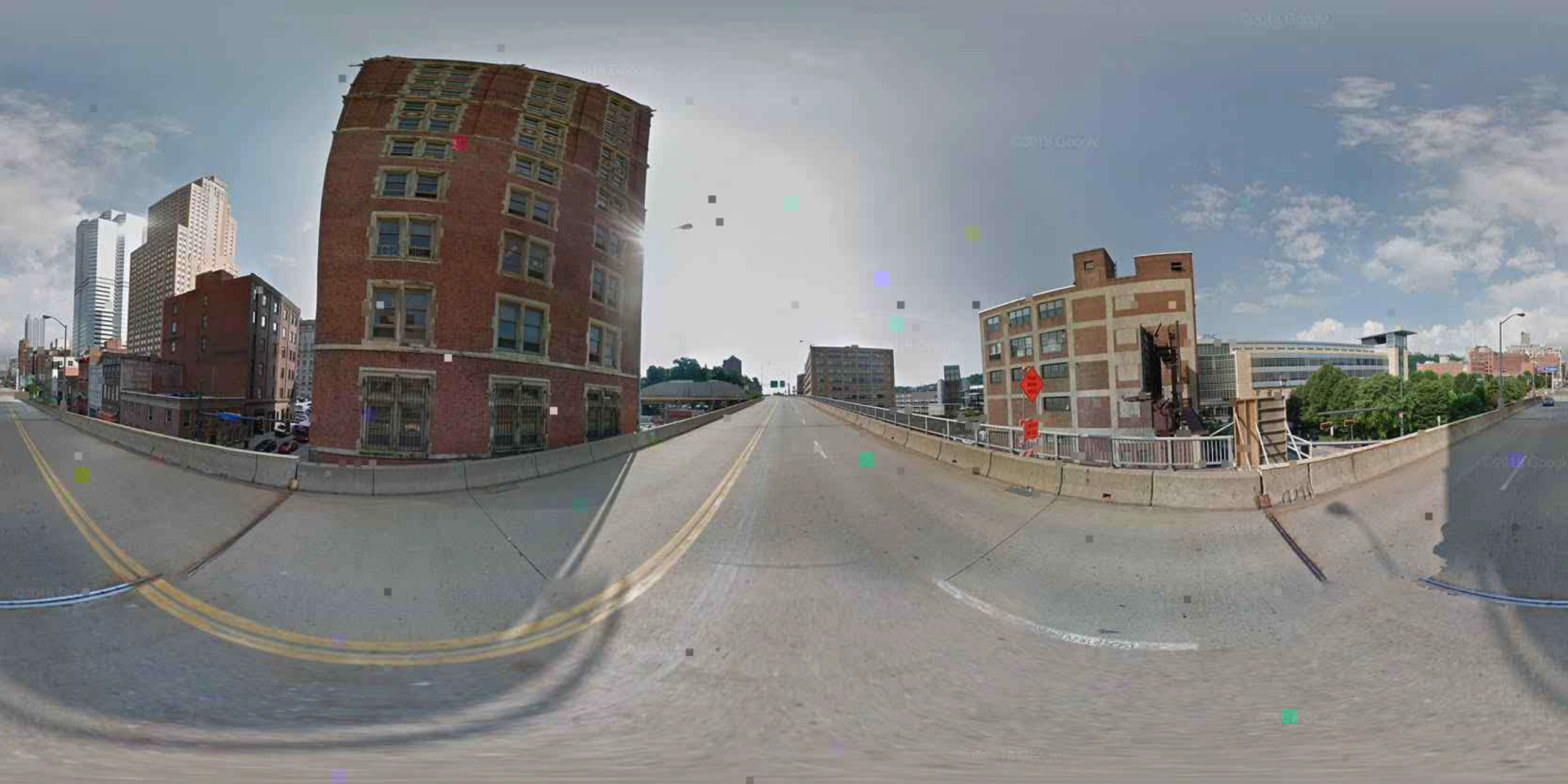}
		\includegraphics[width=0.198\textwidth]{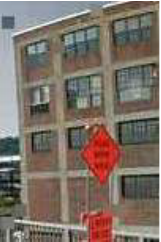}	
		\centering CBR$=0.142$\quad PSNR$=32.34$
		\vspace{0.05cm}
	\end{minipage}
	
	\caption{The decoded images of ParaSC and other baseline schemes under AWGN channels in terms of different SNRs, with their CBR/PSNR(dB) performance at the bottom. A close-up is attached to each image on the right.}
	\label{Visual1}
\end{figure*}
\begin{figure*}[tbp]
	\begin{minipage}[t]{0.11\linewidth}
		\quad
	\end{minipage}
	\hspace{-1ex} {\large Noisy Compressed Image} \hspace{11ex} {\large Weights of $\bm{u}_1$} \hspace{16ex}  {\large Decoded Image} \\
	\begin{minipage}[t]{0.11\linewidth}
		\centering
		\vspace*{-1.4ex} 
		SNR$=2$dB
	\end{minipage}
	\centering
	\begin{minipage}[tbp]{0.29\linewidth}
		\centering
		\subfloat{
			\centering
			\includegraphics[width=1\textwidth]{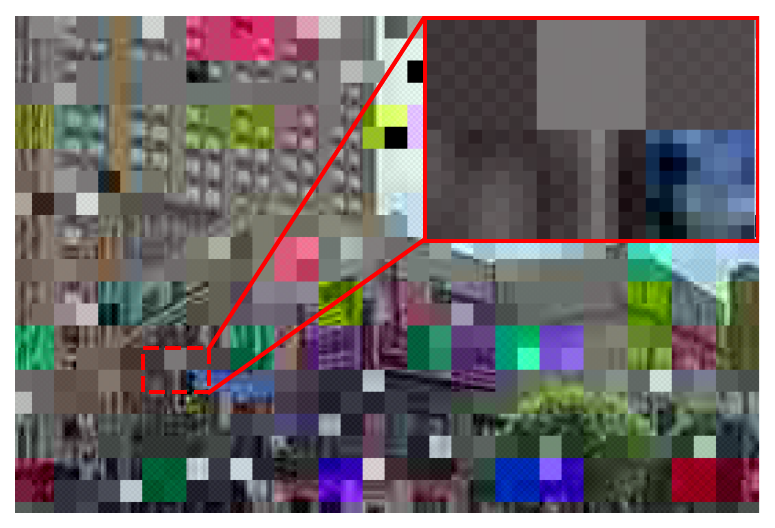}
		}
	\end{minipage}
	\begin{minipage}[tbp]{0.29\linewidth}
		\centering
		\subfloat{
			\centering
			\includegraphics[width=1\textwidth]{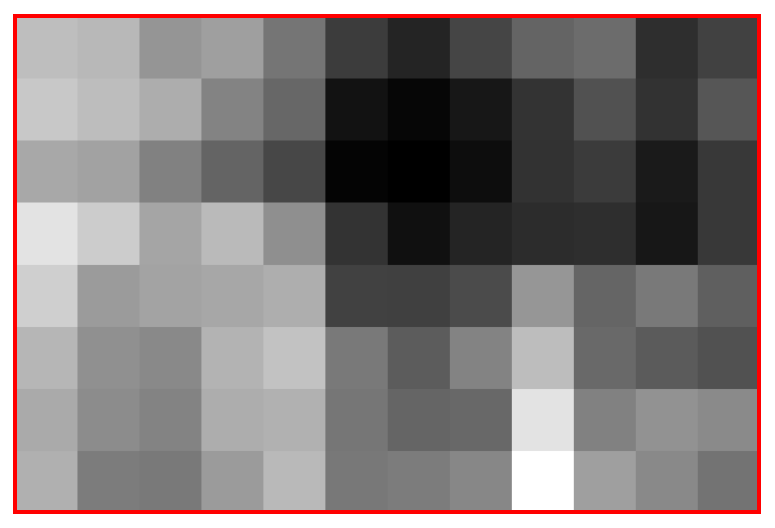}
		}
	\end{minipage}
	\begin{minipage}[tbp]{0.29\linewidth}
		\centering
		\subfloat{
			\centering
			\includegraphics[width=1\textwidth]{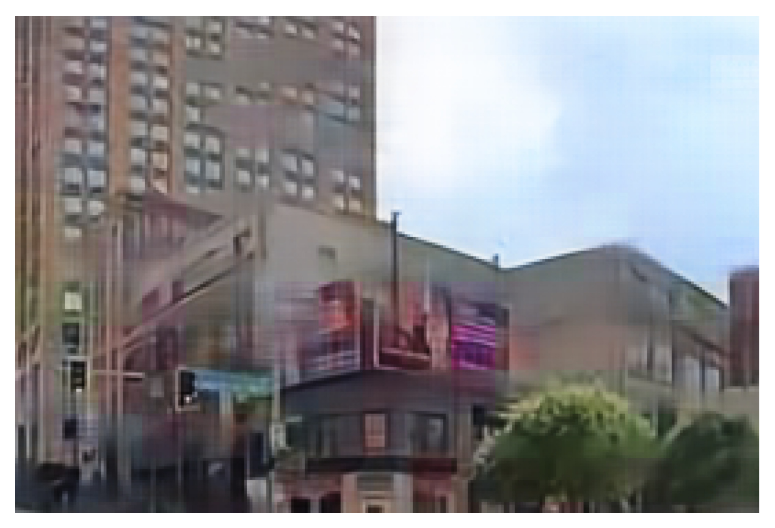}
		}
	\end{minipage}\\
	\begin{minipage}[t]{0.11\linewidth}
		\centering
		\vspace*{-1.5ex} 
		SNR$=10$dB
	\end{minipage}
	\centering
	\begin{minipage}[tbp]{0.29\linewidth}
		\centering
		\subfloat{
			\centering
			\includegraphics[width=1\textwidth]{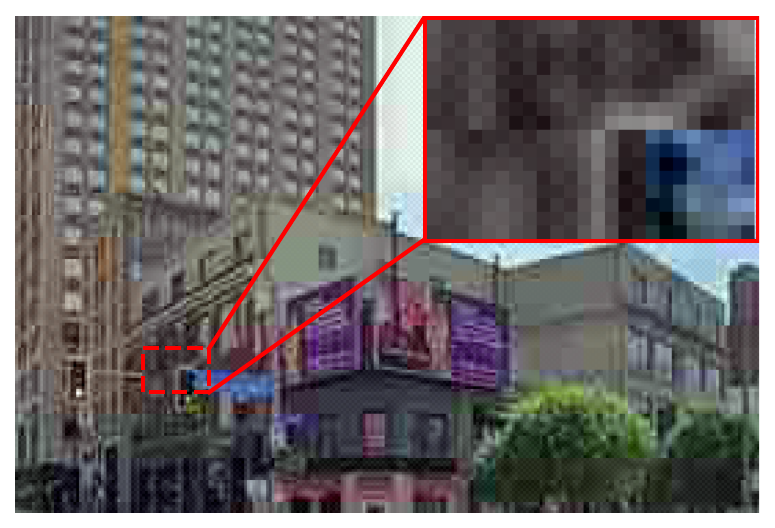}
		}
	\end{minipage}
	\begin{minipage}[tbp]{0.29\linewidth}
		\centering
		\subfloat{
			\centering
			\includegraphics[width=1\textwidth]{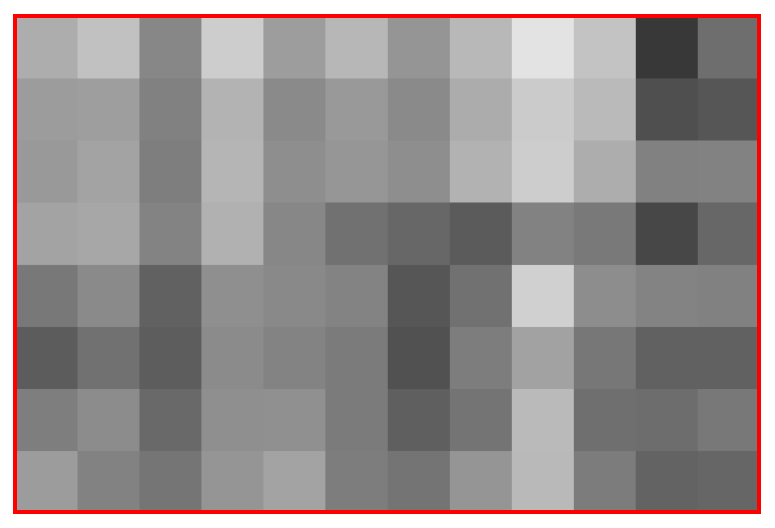}
		}
	\end{minipage}
	\begin{minipage}[tbp]{0.29\linewidth}
		\centering
		\subfloat{
			\centering
			\includegraphics[width=1\textwidth]{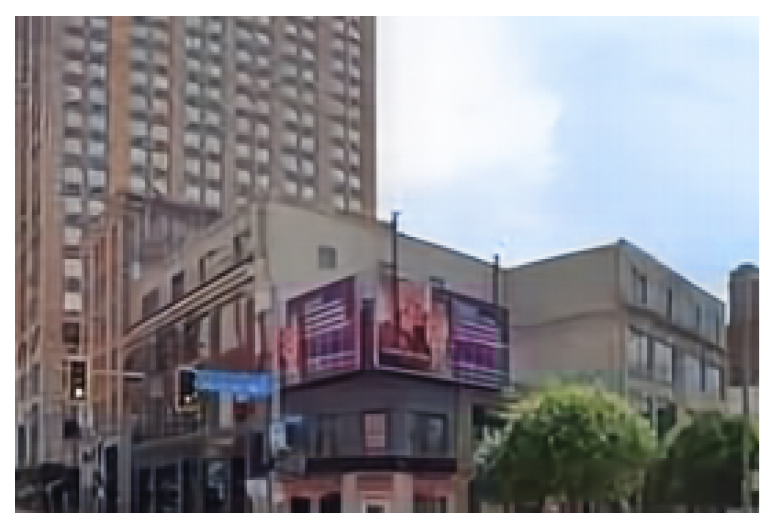}
		}
	\end{minipage}
	\caption{Visualizations of noisy compressed images $\hat{\bm{x}}_c$, aggregation weights of $\bm{u}_1$, and decoded images $\hat{\bm{x}}$ during ParaSC transmission under AWGN channel with varied SNRs. The plotted weights correspond to the region within the red box in $\hat{\bm{x}}_c$. }
	\label{pagnet_visual}
\end{figure*}

\subsection{Complexity Analysis}
In this section, we compare the complexity of ParaSC with state-of-the-art DL-based semantic communication frameworks. Additionally, we involve wireless image transmission transformer (WITT) framework \cite{witt} in comparison. Note that we focus on the transmitter at the distributed edge of the aforementioned frameworks, since many realistic systems mainly suffer from the heavily-restricted resources at the transmitter. 

In Table.~\ref{Tab3}, the number of model parameters, computational complexity, and throughput are presented in terms of four peer frameworks. It can be observed that on every metric, our architecture performs better than all the others except the DJSCC, which adopts a compact auto-encoder structure. Compared with the CNN-based frameworks, WITT appears much more resource-consuming due to the self-attention mechanism of its transformer. 
\subsection{Visualizations of Reconstruction}
In this section, we discuss the visualizations of reconstructed results for AWGN channels in Fig.~\ref{Visual1}. It is observed that for different SNRs, ParaSC reconstructs high-quality images while consuming the fewest bandwidth cost. Particularly, the checkerboard artifact occurs in DJSCC, which mainly results from transpose convolution~\cite{deconv}. Despite the same transpose convolution structure, ParaSC reconstructs the image with smoother texture and higher perceptual quality. In the presence of low channel capacity, the reconstructed image of conventional codec is contaminated with color noises, which are inhibited in ParaSC. Fig.~\ref{Visual1} verifies that the images transmitted by ParaSC are highly aligned with human vision.
\subsection{Robustness to Cliff Effect of SSCC}
Finally, we investigate the effect of PAGNet on overcoming the cliff effect of pure SSCC transmission in Fig.~\ref{pagnet_visual}, where the noisy compressed image $\hat{\bm{x}}_c$, the aggregation weight of the first down-sampling latent $\bm{u}_1$, and the decoded image $\hat{\bm{x}}$ are displayed under clean and noisy channels. Note that $\bm{u}_1$ is obtained from the first down-sampling block in the semantic decoder, thus carrying the information from the image stream. Since the weights are normalized by softmax functions, the sum of the weights for the semantic and image streams at the same pixel is 1, which means a lower pixel weight for the image stream (a deeper color in the second column of Fig.~\ref{pagnet_visual}) equals a higher pixel weight for the semantic stream. To better highlight the role of PAGNet, we extract a small region from $\hat{\bm{x}}_c$ within the red box to visualize the corresponding weights.

It can be observed that in a low-noise channel (SNR$=$$10$dB), $\hat{\bm{x}}_c$ shows a high quality due to the error-free transmission of the image stream. In this case, the aggregation weights of $\bm{u}_1$ are distributed quite evenly across the spatial domain. Conversely, as the channel becomes highly noisy (SNR$=$$2$dB), the image decoded by SSCC is severely degraded by meaningless color patches, especially pure-gray patches. In this case, we note that PAGNet accurately masks these patches with lower weights, while in turn allocating higher weights to these patches in the semantic stream. In this way, ParaSC can successfully inhibit the detrimental effect of channel noises on the final reconstructed image.

\section{Conclusion}\label{Sec6}
In this work, we propose a novel parallel-stream semantic communication framework called ParaSC for image transmission. This framework consists of a conventional image codec, a CNN-based semantic codec and a conditional rate-adapative module. The transmitted semantics are enhanced by the image residual, while the decoder integrates both image stream and semantic stream for reconstruction and assigns SNR-modulated weights to each stream for channel noise resistance. Furthermore, an entropy model based conditional rate-adaptive mechanism is utilized for dynamic conventional coding situations in order to reduce the semantic stream cost. Numerical results show that ParaSC outperforms both conventional image codec and DL-based frameworks in a broad range of scenarios. Moreover, compared to single-stream SSCC methods, our framework brings substantial performance boost in exchange for little extra semantic transmission overhead and low complexity, which supports the compatibility to the current SSCC communication framework.

\bibliographystyle{IEEEtran}
\bibliography{IEEEabrv,Wu_TW-Apr-24-0864}

% Generated by IEEEtran.bst, version: 1.14 (2015/08/26)
\begin{thebibliography}{10}
\providecommand{\url}[1]{#1}
\csname url@samestyle\endcsname
\providecommand{\newblock}{\relax}
\providecommand{\bibinfo}[2]{#2}
\providecommand{\BIBentrySTDinterwordspacing}{\spaceskip=0pt\relax}
\providecommand{\BIBentryALTinterwordstretchfactor}{4}
\providecommand{\BIBentryALTinterwordspacing}{\spaceskip=\fontdimen2\font plus
\BIBentryALTinterwordstretchfactor\fontdimen3\font minus
  \fontdimen4\font\relax}
\providecommand{\BIBforeignlanguage}[2]{{%
\expandafter\ifx\csname l@#1\endcsname\relax
\typeout{** WARNING: IEEEtran.bst: No hyphenation pattern has been}%
\typeout{** loaded for the language `#1'. Using the pattern for}%
\typeout{** the default language instead.}%
\else
\language=\csname l@#1\endcsname
\fi
#2}}
\providecommand{\BIBdecl}{\relax}
\BIBdecl

\bibitem{jpeg}
G.~K. Wallace, ``The {JPEG} still picture compression standard,'' \emph{Commun.
  ACM}, vol.~34, no.~4, pp. 30--44, Apr. 1991.

\bibitem{jpeg2000}
D.~S. Taubman, M.~W. Marcellin, and M.~Rabbani, ``{JPEG}2000: Image compression
  fundamentals, standards and practice,'' \emph{J. Electron. Imaging}, vol.~11,
  no.~2, pp. 286--287, Apr. 2002.

\bibitem{bpg}
\BIBentryALTinterwordspacing
F.~Bellard. {BPG} image format. (2018 Apr 21). [Online]. Available:
  \url{https://bellard.org/bpg/}
\BIBentrySTDinterwordspacing

\bibitem{ldpc}
T.~Richardson and S.~Kudekar, ``Design of low-density parity check codes for
  5{G} new radio,'' \emph{{IEEE} Commun. Mag.}, vol.~56, no.~3, pp. 28--34,
  Mar. 2018.

\bibitem{5075875}
E.~Arikan, ``Channel polarization: A method for constructing capacity-achieving
  codes for symmetric binary-input memoryless channels,'' \emph{{IEEE} Trans.
  Inf. Theory}, vol.~55, no.~7, pp. 3051--3073, Jul. 2009.

\bibitem{Burks_Shannon_Weaver_1951}
C.~E. Shannon, ``A mathematical theory of communication,'' \emph{Bell Syst.
  Tech. J.}, vol.~27, no.~3, pp. 379--423, Jul. 1948.

\bibitem{huang2024compression}
Y.~Huang, J.~Zhang, Z.~Shan, and J.~He, ``Compression represents intelligence
  linearly,'' \emph{arXiv:2404.09937}, 2024.

\bibitem{liu2022time}
C.~Liu, M.~Wang, Z.~Dong, and P.~Wang, ``Time-varying channel estimation based
  on air-ground channel modelling and modulated learning networks,''
  \emph{Chinese J. Electron.}, vol.~31, no.~3, pp. 430--441, May 2022.

\bibitem{10574376}
L.~Wang, L.~Li, L.~Xu, X.~Peng, and A.~Fei, ``Failure-resilient distributed
  inference with model compression over heterogeneous edge devices,''
  \emph{{IEEE} Trans. Mobile Comput.}, vol.~23, no.~12, pp. 12\,680--12\,692,
  Dec. 2024.

\bibitem{9083671}
L.~Wang, J.~Zhang, J.~Chuan, R.~Ma, and A.~Fei, ``Edge intelligence for mission
  cognitive wireless emergency networks,'' \emph{{IEEE} Wireless Commun.},
  vol.~27, no.~4, pp. 103--109, Aug. 2020.

\bibitem{10261454}
L.~Wang, X.~Wu, Y.~Zhang, X.~Zhang, L.~Xu, Z.~Wu, and A.~Fei,
  ``Deep{A}da{I}n-{N}et: Deep adaptive device-edge collaborative inference for
  augmented reality,'' \emph{{IEEE} J. Sel. Topics Signal Process.}, vol.~17,
  no.~5, pp. 1052--1063, Sep. 2023.

\bibitem{Fresia2010}
M.~Fresia, F.~Peréz-Cruz, H.~V. Poor, and S.~Verdú, ``Joint source and
  channel coding,'' \emph{{IEEE} Signal Process. Mag.}, vol.~27, no.~6, pp.
  104--113, Nov. 2010.

\bibitem{10158995}
E.~Erdemir, T.-Y. Tung, P.~L. Dragotti, and D.~Gündüz, ``Generative joint
  source-channel coding for semantic image transmission,'' \emph{{IEEE} J. Sel.
  Areas Commun.}, vol.~41, no.~8, pp. 2645--2657, Aug. 2023.

\bibitem{Wang2023}
S.~Wang, J.~Dai, Z.~Liang, K.~Niu, Z.~Si, C.~Dong, X.~Qin, and P.~Zhang,
  ``Wireless deep video semantic transmission,'' \emph{{IEEE} J. Sel. Areas
  Commun.}, vol.~41, no.~1, pp. 214--229, Jan. 2023.

\bibitem{xie2021}
H.~Xie, Z.~Qin, G.~Y. Li, and B.-H. Juang, ``Deep learning enabled semantic
  communication systems,'' \emph{{IEEE} Trans. Signal Process.}, vol.~69, pp.
  2663--2675, 2021.

\bibitem{xiao2023}
Z.~Xiao, S.~Yao, J.~Dai, S.~Wang, K.~Niu, and P.~Zhang, ``Wireless deep speech
  semantic transmission,'' in \emph{Proc. IEEE Int. Conf. Acoust., Speech
  Signal Process. (ICASSP)}, Rhodes Island, Greece, Jun. 2023, pp. 1--5.

\bibitem{Huang2022}
Y.~{Huang}, B.~{Bai}, Y.~{Zhu}, X.~{Qiao}, X.~{Su}, and P.~{Zhang}, ``{ISC}om:
  Interest-aware semantic communication scheme for point cloud video
  streaming,'' \emph{arXiv:2210.06808}, 2022.

\bibitem{10093953}
B.~Zhao, H.~Xing, X.~Wang, Z.~Xiao, and L.~Xu, ``Classification-oriented
  distributed semantic communication for multivariate time series,''
  \emph{{IEEE} Signal Process. Lett.}, vol.~30, pp. 369--373, 2023.

\bibitem{djscc-l}
D.~B. Kurka and D.~Gündüz, ``Bandwidth-agile image transmission with deep
  joint source-channel coding,'' \emph{{IEEE} Trans. Wireless Commun.},
  vol.~20, no.~12, pp. 8081--8095, Dec. 2021.

\bibitem{dynamicjscc}
M.~Yang and H.-S. Kim, ``Deep joint source-channel coding for wireless image
  transmission with adaptive rate control,'' in \emph{Proc. IEEE Int. Conf.
  Acoust., Speech Signal Process. (ICASSP)}, Singapore, Singapore, May 2022,
  pp. 5193--5197.

\bibitem{ntscc}
J.~Dai, S.~Wang, K.~Tan, Z.~Si, X.~Qin, K.~Niu, and P.~Zhang, ``Nonlinear
  transform source-channel coding for semantic communications,'' \emph{{IEEE}
  J. Sel. Areas Commun.}, vol.~40, no.~8, pp. 2300--2316, Aug. 2022.

\bibitem{witt}
K.~Yang, S.~Wang, J.~Dai, K.~Tan, K.~Niu, and P.~Zhang, ``{WITT}: A wireless
  image transmission transformer for semantic communications,'' in \emph{Proc.
  IEEE Int. Conf. Acoust., Speech Signal Process. (ICASSP)}, Rhodes Island,
  Greece, Jun. 2023, pp. 1--5.

\bibitem{9830752}
H.~Xie, Z.~Qin, X.~Tao, and K.~B. Letaief, ``Task-oriented multi-user semantic
  communications,'' \emph{{IEEE} J. Sel. Areas Commun.}, vol.~40, no.~9, pp.
  2584--2597, Sep. 2022.

\bibitem{transformer}
A.~Vaswani, N.~Shazeer, N.~Parmar, J.~Uszkoreit, L.~Jones, A.~N. Gomez,
  {\L}.~Kaiser, and I.~Polosukhin, ``Attention is all you need,'' in
  \emph{Proc. Advances Neural Inf. Process. Syst.}, vol.~30, 2017.

\bibitem{qin2021semantic}
Z.~Qin, X.~Tao, J.~Lu, W.~Tong, and G.~Y. Li, ``Semantic communications:
  Principles and challenges,'' \emph{arXiv:2201.01389}, 2021.

\bibitem{Huang20212}
D.~Huang, X.~Tao, F.~Gao, and J.~Lu, ``Deep learning-based image semantic
  coding for semantic communications,'' in \emph{IEEE Global Commun. Conf.
  (GLOBECOM)}, Madrid, Spain, Dec. 2021, pp. 1--6.

\bibitem{nguyen2019efficient}
T.~T.~B. Nguyen, T.~Nguyen~Tan, and H.~Lee, ``Efficient {QC-LDPC} encoder for
  5{G} new radio,'' \emph{Electronics}, vol.~8, no.~6, p. 668, Jun. 2019.

\bibitem{unet}
O.~Ronneberger, P.~Fischer, and T.~Brox, ``{U}-{N}et: Convolutional networks
  for biomedical image segmentation,'' in \emph{Proc. Med. Image Comput.
  Comput.-Assisted Intervention (MICCAI)}, Munich, Germany, Oct. 2015, pp.
  234--241.

\bibitem{rrdb}
X.~Wang, K.~Yu, S.~Wu, J.~Gu, Y.~Liu, C.~Dong, Y.~Qiao, and C.~Change~Loy,
  ``{ESRGAN}: Enhanced super-resolution generative adversarial networks,'' in
  \emph{Proc. 15 th Eur. Conf. Comput. Vis. Workshops}, Munich, Germany, Sep.
  2018, pp. 1--17.

\bibitem{gdn}
J.~{Ball{\'e}}, V.~{Laparra}, and E.~P. {Simoncelli}, ``{Density modeling of
  images using a generalized normalization transformation},'' in \emph{Proc.
  Int. Conf. Learn. Represent. (ICLR)}, San Juan, Puerto Rico, May 2016.

\bibitem{Blei2022}
D.~M.~Blei, A.~Kucukelbir, and J.~D.~McAuliffe, ``Variational inference: A
  review for statisticians,'' \emph{J. Amer. Stat. Assoc.}, vol. 112, no. 518,
  pp. 859--877, Jul. 2017.

\bibitem{ballé2017endtoend}
J.~Ballé, V.~Laparra, and E.~P. Simoncelli, ``End-to-end optimized image
  compression,'' in \emph{Proc. Int. Conf. Learn. Represent. (ICLR)}, Toulon,
  France, Apr. 2017.

\bibitem{Li2021}
J.~Li, B.~Li, and Y.~Lu, ``Deep contextual video compression,'' \emph{Proc.
  Advances Neural Inf. Process. Syst.}, vol.~34, pp. 18\,114--18\,125, 2021.

\bibitem{ballé2018variational}
J.~{Ball{\'e}}, D.~{Minnen}, S.~{Singh}, S.~J. {Hwang}, and N.~{Johnston},
  ``{Variational image compression with a scale hyperprior},'' in \emph{Proc.
  Int. Conf. Learn. Represent. (ICLR)}, Vancouver, BC, Canada, Apr.-May 2018.

\bibitem{cvrg-pano}
S.~Orhan and Y.~Bastanlar, ``Semantic segmentation of outdoor panoramic
  images,'' \emph{Signal Image Video Process.}, vol.~16, pp. 643--650, Aug.
  2021.

\bibitem{kodak}
\BIBentryALTinterwordspacing
R.~Franzen. {K}odak24 dataset. (1993). [Online]. Available:
  \url{http://r0k.us/graphics/kodak/}
\BIBentrySTDinterwordspacing

\bibitem{div2k}
E.~Agustsson and R.~Timofte, ``Ntire 2017 challenge on single image
  super-resolution: Dataset and study,'' in \emph{Proc. IEEE Conf. Comput. Vis.
  Pattern Recognit. Workshops}, Honolulu, HI, USA, July 2017, pp. 126--135.

\bibitem{perceptualloss}
J.~Johnson, A.~Alahi, and L.~Fei-Fei, ``Perceptual losses for real-time style
  transfer and super-resolution,'' in \emph{Proc. 14 th Eur. Conf. Comput.
  Vis.}, Amsterdam, The Netherlands, Sep. 2016, pp. 694--711.

\bibitem{vgg16}
K.~Simonyan and A.~Zisserman, ``Very deep convolutional networks for
  large-scale image recognition,'' \emph{arXiv:1409.1556}, 2014.

\bibitem{imagenet}
J.~Deng, W.~Dong, R.~Socher, L.-J. Li, K.~Li, and L.~Fei-Fei, ``Image{N}et: A
  large-scale hierarchical image database,'' in \emph{Proc. Conf. Comput. Vis.
  Pattern Recognit.}, Miami, FL, USA, Jun. 2009, pp. 248--255.

\bibitem{Adam}
D.~P. {Kingma} and J.~{Ba}, ``Adam: A method for stochastic optimization,'' in
  \emph{Proc. Int. Conf. Learn. Represent. (ICLR)}, San Diego, CA, USA, 2015.

\bibitem{liu2015parsenet}
W.~Liu, A.~Rabinovich, and A.~C. Berg, ``Parsenet: Looking wider to see
  better,'' \emph{arXiv:1506.04579}, 2015.

\bibitem{djscc}
E.~Bourtsoulatze, D.~Burth~Kurka, and D.~Gündüz, ``Deep joint source-channel
  coding for wireless image transmission,'' \emph{IEEE Trans. Cogn. Commun.
  Netw.}, vol.~5, no.~3, pp. 567--579, Sep. 2019.

\bibitem{msssim}
Z.~Wang, E.~Simoncelli, and A.~Bovik, ``Multiscale structural similarity for
  image quality assessment,'' in \emph{IEEE Asilomar Conf. Signals, Syst.
  Comput.}, vol.~2, 2003, pp. 1398--1402.

\bibitem{lpips}
R.~Zhang, P.~Isola, A.~A. Efros, E.~Shechtman, and O.~Wang, ``The unreasonable
  effectiveness of deep features as a perceptual metric,'' in \emph{Proc. Conf.
  Comput. Vis. Pattern Recognit.}, Salt Lake City, Utah, USA, Jun. 2018, pp.
  586--595.

\bibitem{deconv}
A.~Odena, V.~Dumoulin, and C.~Olah, ``Deconvolution and checkerboard
  artifacts,'' \emph{Distill}, 2016.

\end{thebibliography}
\end{document}